\documentclass[prl,aps,superscriptaddress,footinbib,twocolumn]{revtex4-2}
\usepackage[latin9]{inputenc}
\usepackage{color}
\usepackage{amsmath}
\usepackage{amssymb}
\usepackage{stmaryrd}
\usepackage{graphicx}
\usepackage[normalem]{ulem}
\usepackage{bm}
\usepackage[unicode=true,bookmarks=true,bookmarksnumbered=false,bookmarksopen=false,breaklinks=false,pdfborder={0 0 1},backref=false,colorlinks=true]
 {hyperref}
\hypersetup{
 linkcolor=magenta, urlcolor=blue, citecolor=blue, pdfstartview={FitH}, hyperfootnotes=false, unicode=true}
 
\usepackage{todonotes}
\makeatletter
\usepackage{amsfonts}\usepackage{tabularx}\usepackage{dcolumn}\usepackage{bm}\usepackage{graphicx}\usepackage{epstopdf}
\usepackage{times}
\usepackage{lineno}
\hypersetup{urlcolor=blue}

\definecolor{jwjcolor}{rgb}{0.1, 0.7, 0.3}

\makeatother

\begin{document}
\nolinenumbers
\title{Long-lived topological time-crystalline order on a quantum processor}

% \affiliation{Department of Physics, ZJU-Hangzhou Global Scientific and Technological Innovation Center, Interdisciplinary Center for Quantum Information, and Zhejiang Province Key Laboratory of Quantum Technology and Device, Zhejiang University, Hangzhou 310027, China\\
% $^2$ Center for Quantum Information, IIIS, Tsinghua University, Beijing 100084, China\\
% $^3$ Alibaba-Zhejiang University Joint Research Institute of Frontier
% Technologies, Hangzhou 310027, China\\
% $^{4}$ Skolkovo Institute of Science and Technology, Moscow 121205, Russia\\
% $^{5}$ Shanghai Qi Zhi Institute, 41th Floor, AI Tower, No. 701 Yunjin Road, Xuhui District, Shanghai 200232, China\\
% $^{6}$ State Key Laboratory of Modern Optical Instrumentation, Zhejiang University, Hangzhou 310027, China\\
% }
\author{Liang Xiang}\thanks{These authors contributed equally}
\affiliation{School of Physics, ZJU-Hangzhou Global Scientific and Technological Innovation Center, and Zhejiang Province Key Laboratory of Quantum Technology and Device, Zhejiang University, Hangzhou, China}

\author{Wenjie Jiang}\thanks{These authors contributed equally}
\affiliation{Center for Quantum Information, IIIS, Tsinghua University, Beijing 100084, China}

\author{Zehang Bao}\thanks{These authors contributed equally}
\author{Zixuan Song}
\author{Shibo Xu}
\author{Ke Wang}
\author{Jiachen Chen}
\author{Feitong Jin}
\author{Xuhao Zhu}
\author{Zitian Zhu}
\author{Fanhao Shen}
\author{Ning Wang}
\author{Chuanyu Zhang}
\author{Yaozu Wu}
\author{Yiren Zou}
\author{Jiarun Zhong}
\author{Zhengyi Cui}
\author{Aosai Zhang}
\author{Ziqi Tan}
\author{Tingting Li}
\author{Yu Gao}
\author{Jinfeng Deng}
\author{Xu Zhang}
\author{Hang Dong}
\author{Pengfei Zhang}
\affiliation{School of Physics, ZJU-Hangzhou Global Scientific and Technological Innovation Center, and Zhejiang Province Key Laboratory of Quantum Technology and Device, Zhejiang University, Hangzhou, China}

\author{Si Jiang}
\author{Weikang Li}
\author{Zhide Lu}
\author{Zheng-Zhi Sun}
\affiliation{Center for Quantum Information, IIIS, Tsinghua University, Beijing 100084, China}

\author{Hekang Li}
\affiliation{School of Physics, ZJU-Hangzhou Global Scientific and Technological Innovation Center, and Zhejiang Province Key Laboratory of Quantum Technology and Device, Zhejiang University, Hangzhou, China}
\author{Zhen Wang}
\affiliation{School of Physics, ZJU-Hangzhou Global Scientific and Technological Innovation Center, and Zhejiang Province Key Laboratory of Quantum Technology and Device, Zhejiang University, Hangzhou, China}
\affiliation{Hefei National Laboratory, Hefei 230088, China}
\author{Chao Song}
\affiliation{School of Physics, ZJU-Hangzhou Global Scientific and Technological Innovation Center, and Zhejiang Province Key Laboratory of Quantum Technology and Device, Zhejiang University, Hangzhou, China}
\author{Qiujiang Guo}
\email{qguo@zju.edu.cn}
\affiliation{School of Physics, ZJU-Hangzhou Global Scientific and Technological Innovation Center, and Zhejiang Province Key Laboratory of Quantum Technology and Device, Zhejiang University, Hangzhou, China}
\affiliation{Hefei National Laboratory, Hefei 230088, China}

\author{Fangli Liu}
\affiliation{Joint Quantum Institute and Joint Center for Quantum Information and Computer Science, NIST and University of Maryland, College Park, MD, USA}
\affiliation{QuEra Computing Inc., Boston, MA, USA}

\author{Zhe-Xuan Gong}
\affiliation{Department of Physics, Colorado School of Mines, Golden, CO, USA}
\affiliation{National Institute of standards and Technology, Boulder, CO, USA}

\author{Alexey V. Gorshkov}
\affiliation{Joint Quantum Institute and Joint Center for Quantum Information and Computer Science, NIST and University of Maryland, College Park, MD, USA}

\author{Norman Y. Yao}
\affiliation{Department of Physics, Harvard University, Cambridge 02138 MA, USA}

\author{Thomas Iadecola}
\affiliation{Department of Physics and Astronomy, Iowa State University, Ames, IA, USA}
\affiliation{Ames Laboratory, Ames, IA, USA}

\author{Francisco Machado}
\affiliation{Department of Physics, Harvard University, Cambridge 02138 MA, USA}
\affiliation{ITAMP, Harvard-Smithsonian Center for Astrophysics, Cambridge, Massachusetts, 02138, USA}

\author{H. Wang}
\email{hhwang@zju.edu.cn}
\affiliation{School of Physics, ZJU-Hangzhou Global Scientific and Technological Innovation Center, and Zhejiang Province Key Laboratory of Quantum Technology and Device, Zhejiang University, Hangzhou, China}
\affiliation{Hefei National Laboratory, Hefei 230088, China}

\author{Dong-Ling Deng}
\email{dldeng@tsinghua.edu.cn}
\affiliation{Center for Quantum Information, IIIS, Tsinghua University, Beijing 100084, China}
\affiliation{Hefei National Laboratory, Hefei 230088, China}
\affiliation{Shanghai Qi Zhi Institute, 41th Floor, AI Tower, No. 701 Yunjin Road, Xuhui District, Shanghai 200232, China}

\begin{abstract} % for a Science paper, the abstract is limitted to 125 words
\textbf{
Topologically ordered phases of matter~\cite{Tsui1982TwoDim,Wen1990Topological,nayak2008non} elude Landau's symmetry-breaking theory~\cite{Landau2013Statistical}, featuring a variety of intriguing properties such as long-range entanglement and intrinsic robustness against local perturbations. 
Their extension to periodically driven systems gives rise to exotic new phenomena that are forbidden in thermal equilibrium~\cite{Khemani2016Phase,Potter2016Classification}. Here, we report the observation of signatures of such a phenomenon---a prethermal topologically ordered time crystal---with programmable superconducting qubits arranged on a square lattice.
By periodically driving the superconducting qubits with a surface-code Hamiltonian~\cite{Kitaev2003Faulttolerantb,Wen2003Quantum}, we observe discrete time-translation symmetry breaking dynamics~\cite{Wilczek2012PRL,Else2016PRL,Yao2017PRL,Khemani2019brief,Zhang2022Digital,Bomantara2021Topological,Chew2020TimeCrystalline,Giergiel2019Topological} that is only manifested in the subharmonic temporal response of nonlocal logical operators. 
We further connect the observed dynamics to the underlying topological order by measuring a nonzero topological entanglement entropy~\cite{Kitaev2006Topological, Levin2006Detecting} and studying its subsequent dynamics.
Our results demonstrate the potential to explore exotic topologically ordered nonequilibrium phases of matter with noisy intermediate-scale quantum processors~\cite{Preskill2018Quantum}.
}
\end{abstract}

\maketitle

Phases of matter are often classified by broken symmetries and local order parameters~\cite{Landau2013Statistical}. However, the discovery of topological order has transformed this simple paradigm~\cite{Tsui1982TwoDim,Wen1990Topological}. Two topologically ordered phases with the same symmetries can showcase topologically distinct features, such as different patterns of long-range entanglement and the emergence of quasiparticles with different anyonic braiding statistics~\cite{nayak2008non,Andersen2023NonAbelian,Xu2023Digital}. These features are intrinsically nonlocal in that they cannot be distinguished by any local order parameter \cite{Kitaev2003Faulttolerantb,Freedman1998NP}. 
Unfortunately, topological order is usually restricted to the ground state; mobile thermal excitations can hybridize nominally degenerate ground states by traversing the system along nontrivial closed loops.
By introducing disorder, the motion of these excitations can be arrested and the hybridization process suppressed. In the limit where excitations are fully localized, the topological phase becomes stable across the entire energy spectrum of the system
\cite{Nandkishore2015many,Abanin2019RMP, Wahl2020Local,Wahl2022Local,Huse2013Localization,Wahl2020Local,Khemani2016Phase,Potter2016Classification,Bauer2013Area,Bahri2015Localization,Huse2013Localization}.

Time-periodic driving of a quantum many-body system enables novel phases of matter that cannot exist in thermal equilibrium. A prominent example is that of time crystals ~\cite{Wilczek2012PRL,Else2016PRL,Yao2017PRL,Khemani2019brief,Zhang2022Digital,Bomantara2021Topological,Chew2020TimeCrystalline,Giergiel2019Topological}, where discrete time translation symmetry is spontaneously broken. Strikingly, the concept of a time crystal can be extended to include topological order, resulting in a new dynamical phase dubbed a topologically ordered time crystal~\cite{Wahl2021arxivTTC}. Unlike conventional time crystals, where the breaking of time translation symmetry manifests in the dynamics of local observables, topologically ordered time crystals show such symmetry breaking only for nonlocal logical operators. 
Whether or not this phase has a truly infinite lifetime depends on the late-time stability of many-body localization~\cite{Doggen2020Slow,Potirniche2019Exploration,DeRoeck2017Manybody}; nevertheless, the dynamical features of the system can still exhibit very long-lived signatures of localization persisting beyond current experimental timescales.
While signatures of conventional time crystals without topological order have been observed in a number of distinct systems, including trapped ions~\cite{Zhang2017Observation,Kyprianidis2021Observation}, spins in nitrogen-vacancy centers~\cite{Choi2017Observation,Randall2021Manybodylocalized}, ultracold atoms~\cite{Smits2018Observation,Autti2018Observation}, solid-state spin ensembles~\cite{OSullivan2020Signatures,Pal2018Temporal,Rovny2018Observation}, and superconducting qubits~\cite{Mi2022Timecrystalline,Ying2022Floquet,Zhang2022Digital}, the observation of a topologically ordered time crystal remains an open challenge. %Realizing 

\begin{figure*}[tb]
\center
\includegraphics[width=1.0\linewidth]{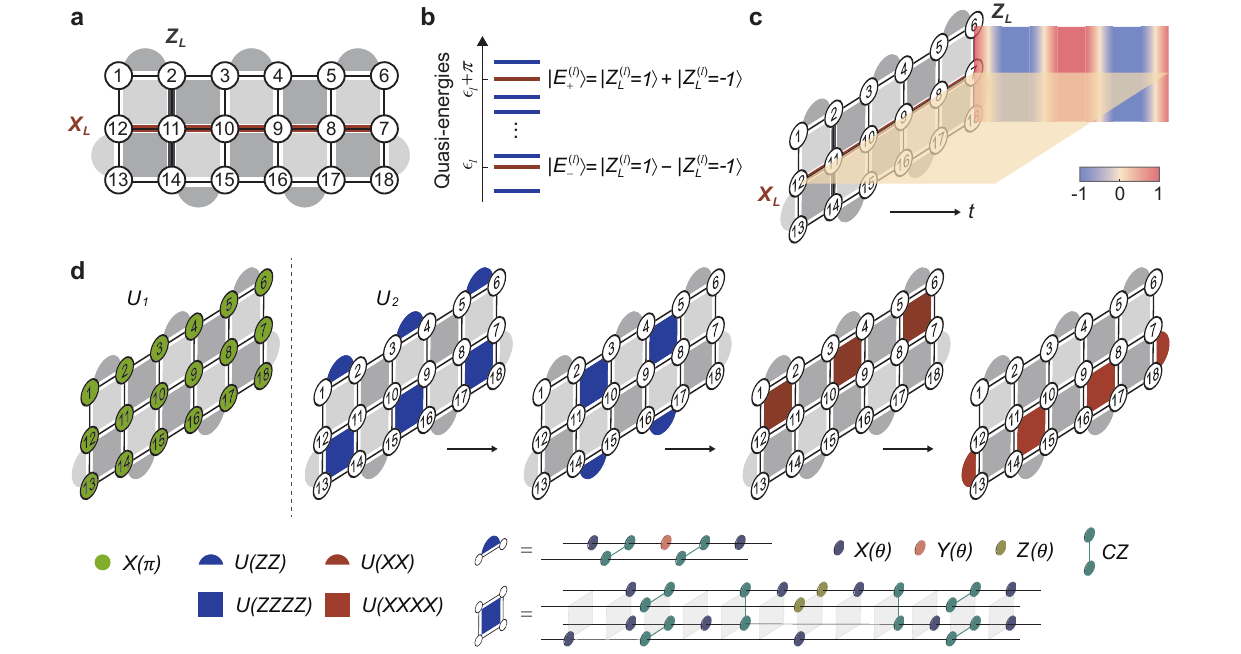}
\caption{\textbf{Periodically driven surface code model.} 
{\bf a}, Rotated surface code model on a three-by-six square lattice. The circled numbers label the qubits. The dark and light gray regions represent plaquette operators ${A}_p$ and ${B}_q$, respectively.  The thick black (red) line represents the nonlocal string operator $Z_L$ ($X_L$).
{\bf b}, Topologically ordered Floquet eigenstates in the limit $B \rightarrow 0$.  The quasi-energies of each pair of eigenstates $|E_{\pm}^{(l)}\rangle$ are split by $\pi$. 
{\bf c}, Schematic of the stroboscopic dynamics of the string operators $Z_L$ and $X_L$. Under periodic driving, the expectation value of $Z_L$  exhibits a persistent subharmonic oscillation with a period of $2T$, while $X_L$ preserves a constant value of zero. 
{\bf d}, Decomposition of the Floquet unitary $U_F$~($B=0$) into elementary quantum gates. $U_1$ is realized by applying $\pi$ pulses to all the qubits. Since all the plaquette operators commute with each other, $U_2$ is constructed by sequentially applying four groups of them. Plaquette unitaries $e^{-i{A}_p{T/2}}$, labeled by $U(ZZZZ)$ and $U(ZZ)$, and  $e^{-i{B}_q{T/2}}$, labeled by $U(XXXX)$ and  $U(XX)$, are further decomposed into sequences of single-qubit rotations and two-qubit controlled-Z gates. $X(\theta)$, $Y(\theta)$, and $Z(\theta)$ denote single-qubit rotations by an angle $\theta$ around the $x$-, $y$-, and $z$-axis, respectively. $e^{-i{B}_q{T/2}}$ can be implemented by sandwiching $e^{-i{A}_q{T/2}}$ with Hadamard gates.  In the experiment, the whole circuit is further compiled  to reduce the depth and suppress hardware noise (Supplementary Section II.E).
}
\label{fig1:model}
\end{figure*}

Here, we report the observation of a long-lived prethermal topologically ordered discrete time crystal, with eighteen programmable superconducting transmon qubits arranged on a two-dimensional square lattice. By optimizing the device fabrication and control process, we push the median lifetime of these qubits to $T_1\approx163~\mu$s and the median simultaneous single- and two-qubit gate fidelities above $99.9\%$ and $99.4\%$, respectively. 
Together with a neuroevolution algorithm \cite{Lu2021Markovian} that outputs near-optimal quantum circuits for digitally simulating four-body interactions, this enables us to successfully implement Floquet surface-code dynamics with an optimized quantum circuit of depth exceeding $700$, consisting of more than $2300$ single- and $1400$ two-qubit gates.  We measure the dynamics of nonlocal logical operators and local spin magnetizations and find that the former show a robust subharmonic response, whereas the latter decay quickly to zero and do not show period-doubled oscillations. This differs drastically from conventional discrete time crystals, where local, rather than nonlocal, observables exhibit subharmonic response. We further reveal the long-range quantum entangled nature of topological order by preparing a many-body eigenstate of the Floquet unitary and measuring its topological entanglement entropy with different subsystem sizes and geometries ~\cite{Kitaev2006Topological, Levin2006Detecting}. We obtain near-expected values for the measured topological entanglement entropy, which deviates significantly from the trivial-state value of zero and provides strong evidence for the presence of topological order. 

\vspace{.5cm}
\begin{figure*}[hbt]
\center
\includegraphics[width=1.0\linewidth]{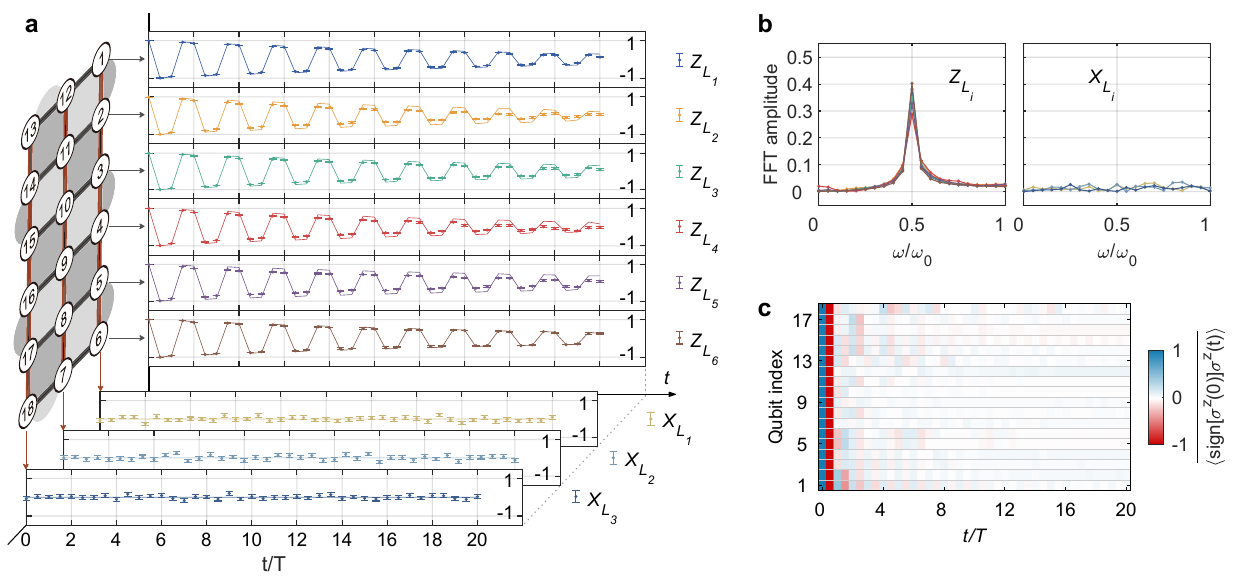}
\caption{\textbf{Time-translation symmetry breaking for nonlocal observables with $B=0$.}
{\bf a}, Dynamics of nonlocal observables.  Auto-correlation function for the three-body string operators $\{Z_{L_i}\}$ (thick black lines) and the six-body string operators  $\{X_{L_i}\}$ (thick red lines) are shown in the upper six and lower three panels, respectively. %\del{Error bars represent the standard error of the statistical mean over 24 random realizations.} 
Experimental data points (dots) are obtained from averaging over 24 random realizations, with error bars representing the standard error of the statistical mean. The numerical results (lines) are computed by taking into account qubit decoherence and gate errors (Supplementary Section III).
Whereas the expectation values for $\{X_{L_i}\}$ remain zero, the auto-correlators for $\{Z_{L_i}\}$ exhibit stable subharmonic oscillations for up to
20 cycles.
{\bf b}, Fourier spectra of time-domain signals observed in {\bf a}, where a stable subharmonic frequency peak appears for $\{Z_{L_i}\}$ but not $\{X_{L_i}\}$. 
{\bf c}, Dynamics of the auto-correlation function for local observables $\{\sigma_k^z\}$. Such auto-correlations decay quickly to zero, in sharp contrast to those of the string operators $\{Z_{L_i}\}$.}
\label{fig2}
\end{figure*}
\vspace{.5cm}
%\noindent\textbf{\large{}Framework and experimental setup}{\large\par}

\begin{figure}
\includegraphics[width=1.0\linewidth]{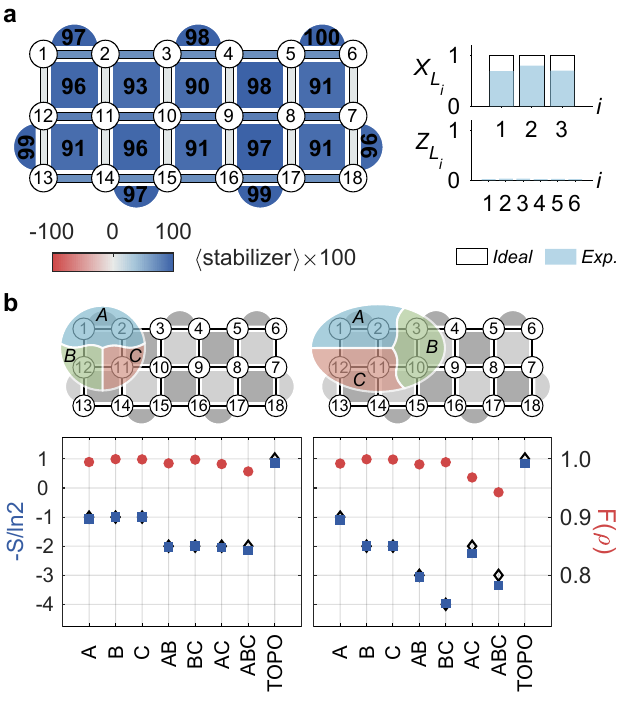} 
\caption{\textbf{Measured stabilizer values and topological entanglement entropy for a Floquet eigenstate.} 
{\bf a}, Measured expectation values of plaquette operators $\{A_p\}$ and $\{B_q\}$ for the Floquet eigenstate $|E^{(0)}_+\rangle$  are shown in the left panel.  Expectation values of string operators $\{X_{L_i}, Z_{L_i}\}$ are plotted in the right panel with solid (hollow) bars representing experimental (theoretical) results. 
{\bf b}, Measuring topological entanglement entropy (TOPO, $S_{\rm topo}$) on four- and six-qubit subsystems. Division of subsystems and corresponding experimental results are shown in the upper and lower panels, respectively.  Blue squares (red circles) represent entanglement entropy (region fidelities $F(\rho_i)$). The topological entanglement entropy is extracted from the measured von Neumann entropy for the regions $S_i$. The error bars, obtained by repeated measurements, are very small and not shown here. The black rhombus markers are theoretical predictions for $S_{\rm topo}$ and $S_i$, which agree well with the corresponding experimental results.}
\label{fig3}
\end{figure}

\begin{figure*}[tb]
\center
\includegraphics[width=1.0\linewidth]{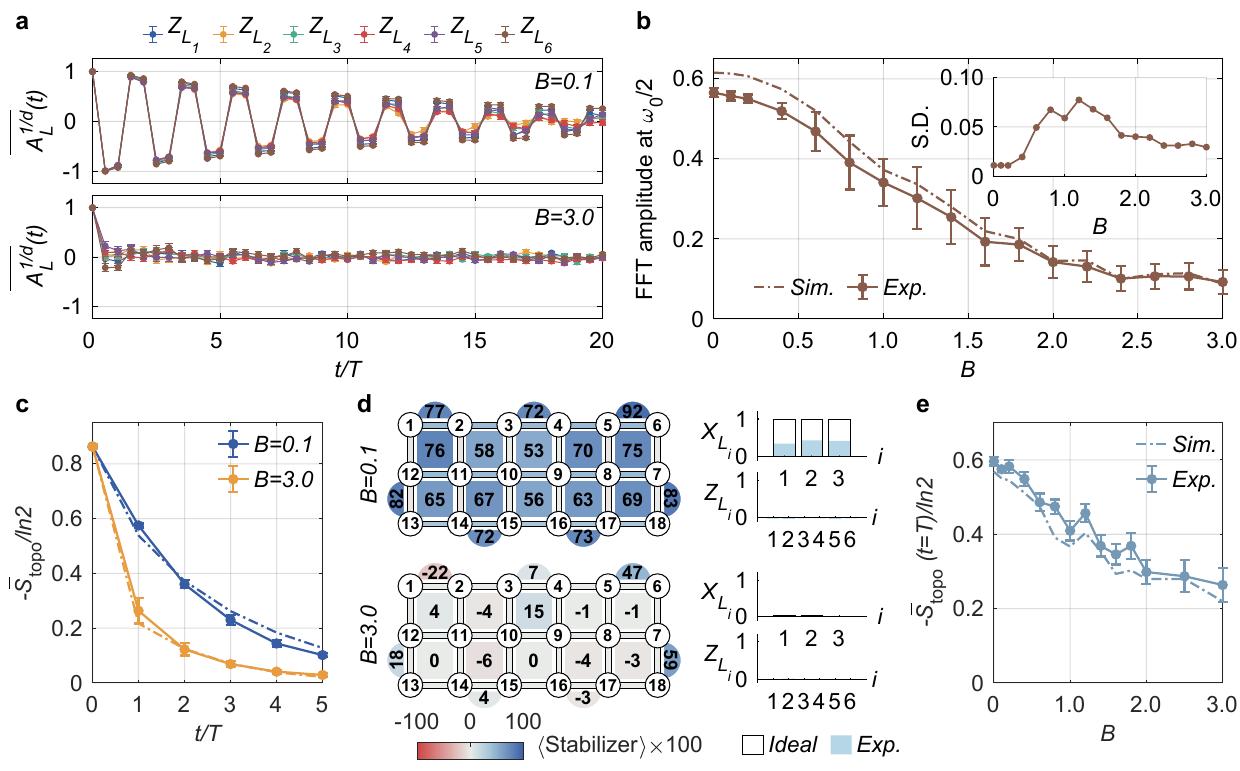}
\vspace{-0.6cm}
\caption{\textbf{Robustness of the topological time-crystalline eigenstate order.}
{\bf a}, Measured disorder-averaged auto-correlation function $\overline{A^{1/d}_{L_i}}(t)$  for string operators $\{Z_{L_i}\}$ with $B=0.1$ (upper panel) and $B=3.0$~(lower panel).  Error bars denote the standard error of the statistical mean over 24 random realizations.
{\bf b}, Amplitudes of Fourier spectra at $\omega/\omega_0=0.5$ as a function of $B$.   Fourier transform of $A^{1/d}_{L}(t)$ is performed using averaged time-domain signals over $\{Z_{L_i}\}$ for up to $t=6T$. Each data point is averaged over 24 random realizations. Error bars are the standard deviation (S.D.) for 24 disorder realizations. Insert: The S.D. of Fourier spectra amplitudes at $\omega/\omega_0=0.5$ as a function of $B$.
%\del{Sample-to-sample fluctuations, shown in the inset, are obtained by calculating the standard deviation (S.D.) of 24 disorder realizations for each  $Z_{L_i}$ and then averaging over  $\{Z_{L_i}\}$. Error bars are the standard deviation over $\{Z_{L_i}\}$.}
{\bf c}, Quench dynamics of the disorder-averaged topological entanglement entropy $\overline{S}_{\rm topo}$ from the initial state $|E^{(0)}_+\rangle$ (which is a Floquet eigenstate at $B=0$) for different $B$. Here, $\overline{S}_{\rm topo}(t \neq0)$ is obtained by performing state tomography on a four-qubit subsystem and the average is over 12 random realizations; $\overline{S}_{\rm topo}(t = 0)$ is obtained via the same state tomography process and averaging over five repetitions of eigenstate preparation. % Since quantum state tomography is time-consuming,  each  data point is averaged over 12 random realizations. Error bars are the standard error of statistical mean.
{\bf d}, Measured plaquette and string operators for the eigenstate $|E^{(0)}_+\rangle$  after single-step $U_F$ evolution at $B=0.1$ and $B=3.0$. 
{\bf e}, $\overline{S}_{\rm topo}(t=T)$  as a function of random field strength $B$, which is averaged over 12 random realizations. %Error bars are the standard error of statistical mean.
Numerical simulations (dashed lines) in {\bf b}, {\bf c}, and {\bf e} are carried out with noisy quantum gates (see Supplementary Section III for details).
}
\label{fig4}
\end{figure*}

\noindent\textbf{\large{}Theoretical model and experimental setup}{\large\par}
\noindent We consider the periodically driven rotated surface code model on a 2D lattice with open boundary conditions~\cite{Wahl2021arxivTTC, Bomantara2021PRBnonlocal}: 
% \begin{eqnarray}\label{eqs:Floquet_model}
% H(t)&=&\left \{\begin{array}{ll}
%     H_1, & 0 \leq t<T^{\prime}, \\
%     H_2, & T^{\prime} \leq t<T,
%     \end{array} \right \nonumber \\
%     H_1&\equiv& \frac{\pi}{2} \sum_k \sigma_k^x+\sum_k \boldsymbol{B}_k \cdot \boldsymbol{\sigma}_k, \\
%     H_2&\equiv& -\sum_p \alpha_p A_p-\sum_q \beta_q B_q, \nonumber 
% \end{eqnarray}

\begin{eqnarray}\label{eqs:Floquet_model}
\begin{aligned}
H(t) & = \begin{cases}
H_1, & 0 \leq t<T^{\prime}, \\
H_2, & T^{\prime} \leq t<T, 
\end{cases} \\
H_1 & \equiv \frac{\pi}{2} \sum_k \sigma_k^x+\sum_k \boldsymbol{B}_k \cdot \boldsymbol{\sigma}_k, \\
H_2 & \equiv-\sum_p \alpha_p A_p-\sum_q \beta_q B_q,
\end{aligned}
\end{eqnarray}
%\begin{equation}
%\label{eqs:Floquet_model}
%    H(t)=\left\{\begin{array}{ll}
%    H_1=\frac{\pi}{2} \sum_k \sigma_k^x+\sum_k \boldsymbol{B}_k \cdot \boldsymbol{\sigma}_k, & 0 \leq t<T^{\prime} \\
%    H_2=-\sum_p \alpha_p A_p-\sum_q \beta_q B_q, & T^{\prime} \leq t<T
%    \end{array}\right.
%\end{equation}
where $\boldsymbol{{\sigma}}_k=({\sigma}_k^{x},{\sigma}_k^{y},{\sigma}_k^{z})$ is a vector of Pauli matrices acting on the $k$-th qubit; $\boldsymbol{B}_k$ denotes an on-site field drawn randomly and independently  from a ball with radius $B$;  %random vector chosen in a sphere with radius $B$;
the plaquette operators ${A}_p=\prod_{m\in p}{\sigma}_m^z$ and ${B}_q=\prod_{n\in q}{\sigma}_n^x$ are products of Pauli operators on the corresponding plaquettes (Fig.~\ref{fig1:model}{a}); $\alpha_p$ and $\beta_q$ are coefficients uniformly chosen from $[0,2\pi)$; the drive period is fixed as $T=2T'=2$, which roughly corresponds to a $1.4$-$\mu$s runtime for the corresponding quantum circuit in our experiment. 

We note that, other than the discrete time-translation symmetry, $H(t)$ breaks all microscopic symmetries due to the presence of the random on-site fields $\bm{B}_k$ in $H_1$. The Floquet unitary that fully characterizes the dynamics of the system reads $U_F=U_2U_1$, with $U_1=e^{-iH_1}$ and $U_2=e^{-iH_2}$ being the unitary operators generated by the Hamiltonians $H_1$ and $H_2$, respectively. 
$H_2$ represents the Hamiltonian of the rotated surface code model, whose energy spectrum is two-fold degenerate and whose eigenstates show topological order \cite{Wen2003Quantum,Kitaev2003Faulttolerantb}. Owing to their topological nature, the degenerate eigenstates can only be distinguished by nonlocal string operators such as $Z_L=\prod_{k\in P_z}{\sigma}_k^z$  or $X_L=\prod_{k\in P_x}{\sigma}_k^x$, which traverse the lattice through the path $P_z$ or $P_x$ (see Fig.~\ref{fig1:model}{a}). We label each eigenstate pair by $|Z_L^{(l)}=\pm1\rangle$ for each eigenstate with quasi-energy $\epsilon_l$~(see Supplementary Section I.A). In the limit $B\rightarrow 0$, $U_1$ represents a perfect flip of all spins. As a result, the drive $H_1$ reorganizes the topologically ordered eigenstate pairs of $H_2$ into Floquet eigenstates $|E_{\pm}^{(l)}\rangle$ of the form $|E_{\pm}^{(l)}\rangle \propto |Z_L^{(l)}=1\rangle\pm|Z_L^{(l)}=-1\rangle$. The quasi-energies of the corresponding cat-like eigenstates are split by quasi-energy $\pi$  (Fig.~\ref{fig1:model}{b}). 
As a result,  the stroboscopic dynamics of the nonlocal operator $Z_L$ exhibits a stable subharmonic oscillation with $2T$ periodicity as illustrated in Fig.~\ref{fig1:model}{c}, which breaks the discrete time-translation symmetry by the drive period $T$ (Supplementary Section I.D). These Floquet eigenstates also exhibit topological order, which is essential for the robustness of the subharmonic response of the nonlocal string operators $Z_L$. 

For small but finite $B$, the system's integrability is broken and the eigenstate pairs are no longer exactly split by $\pi$.
However, this deviation arises from the motion of excitations across the system which mixes the different topological sectors, which is strongly suppressed by the disorder in $\alpha_p$ and $\beta_p$.
Until this thermalization occurs, $t\lesssim t_{\mathrm{th}}$, the system's dynamics will exhibit robust period doubling dynamics, much like in the $B=0$ case.
% system's thermalization timescale $t_\text{th}$ remains much larger than experimentally accessible timescales. %the coherence timescale of current quantum devices.
All our experimental and theoretical observations of period doubling behavior pertain to this ``prethermal'' regime which, in the small-$B$ regime, is much larger than the experimentally accessible timescales.

Our experiments are carried out on a programmable flip-chip superconducting processor with $18$ transmon qubits arranged on a 2D square lattice (see Supplementary Section II.A for detailed information about the device). To implement $H(t)$, the four-body terms with random strengths in $H_2$, which are vital for the eigenstate topological order at high energy, pose an apparent challenge since four-body interactions do not naturally appear in the superconducting system. We therefore exploit the idea of digital quantum simulation
to implement $H(t)$ with quantum circuits (Fig.~\ref{fig1:model}{d}), which are obtained via a neuroevolution algorithm \cite{Lu2021Markovian} (Supplementary Section I.E). We mention that these quantum circuits are near-optimal and can implement $H(t)$ in an analytical fashion without any Trotter error, independent of $\alpha_p$, $\beta_p$, and $\bm{B}_k$. With these efficient quantum circuits, improved gate fidelities, and coherence times, we are able to implement and probe the unconventional dynamics of the system up to 20 driving periods.

\vspace{.5cm}
\noindent \textbf{\large{}Subharmonic response for nonlocal observables}{\large\par}

\noindent 
The characteristic signature of topological time-crystalline eigenstate order is the breaking of the discrete time-translation symmetry for nonlocal logical operators, manifested by persistent oscillations with period $2T$. To this end, we define the normalized auto-correlation function  $A^{1/d}_{L}(t)={\rm sign}[\langle Z_{L}(0)Z_{L}(t)\rangle]|\langle Z_{L}(t)\rangle|^{1/d}$  for the $d$-body string operator $Z_L$, where  $\langle\cdots\rangle$ represents the expectation value and the $d$-th root is used to indicate the geometric mean value. We begin by studying the evolution of the disorder-averaged auto-correlator $\overline{A^{1/d}_{L_i}}(t)$  for operators $\{Z_{L_i}\}$ ($d=3$) at the solvable limit $B=0$,
which are averaged over 24 random realizations by sampling Hamiltonian parameters $\alpha_p$,  $\beta_q$ and initial product states.  From Fig.~\ref{fig2}{a}, it is evident that, in the topologically ordered regime, $\overline{A^{1/d}_{L_i}}(t)$ oscillates with a $2T$ periodicity for up to $20$ driving cycles. We mention that $\overline{A^{1/d}_{L_i}}(t)$ exhibits a gradually decaying envelope due to extrinsic experimental imperfections, rather than internal thermalization, which is confirmed by numerical simulations (lines in Fig.~\ref{fig2}{a}) incorporating experimentally measured gate errors and decoherence times. Indeed, the ideal numerical simulations show that the internal thermalization time of the system without experimental noise is far longer than $20$ driving cycles (Supplementary Section I.C). In the frequency domain, $\overline{A^{1/d}_{L_i}}$ shows a peak at the subharmonic frequency of the drive period $\omega/\omega_0=0.5$, as shown in Fig.~\ref{fig2}{b}. We also note that the string operator $X_L$ does not show period-doubled oscillations, and no subharmonic peak is observed in the frequency domain.

Although the $2T$-period subharmonic oscillations of  nonlocal observables $\{Z_{L_i}\}$ already sharply distinguish our experiment from previous works \cite{Zhang2017Observation, Choi2017Observation,Kyprianidis2021Observation, Mi2022Timecrystalline,Zhang2022Digital}, where only local observables break time-translation symmetry, we further demonstrate that the observed Floquet topological order is a nonlocal effect by contrasting with the dynamical behavior of local operators $\{\sigma_k^z\}$.  The auto-correlation function $\overline{\langle {\rm sign}[\sigma_k^z(0)] \sigma_k^z(t) \rangle}$ decays to zero quickly without evident oscillations (Fig.~\ref{fig2}{c}), even though the periodic drive is locally applied to each qubit.  The striking contrast between nonlocal operators $\{Z_{L_i}\}$ and local operators $\{\sigma_k^z\}$  exposes the locally indistinguishable nature of the Floquet topological order and rules out the possibility of trivial oscillations arising from driving a noninteracting system.

\vspace{.5cm}
\noindent \textbf{\large{}Topologically-ordered Floquet eigenstates}{\large\par}

\noindent The Floquet eigenstates bear intrinsic topological order and exhibit long-range quantum entanglement characterized by the topological entanglement entropy $S_{\rm topo}$~\cite{Kitaev2006Topological, Levin2006Detecting} (see Supplementary Section I.B). To reveal the underlying global entanglement, we prepare an eigenstate of $U_F$ and measure its $S_{\rm topo}$ for different system sizes.
In the $B \rightarrow 0$ limit, eigenstates of $U_F$ correspond to superpositions of  degenerate eigenstates of ${H}_2$ (see Fig.~\ref{fig1:model}b). The eigenstate we prepare is the symmetric superposition of ground states of $H_2$, given by $|E^{(0)}_+\rangle=\frac{1}{\sqrt{2}}(|Z_L^{(0)}=1\rangle + |Z_L^{(0)}=-1\rangle)$. 
We prepare it from a simple initial product state using a quantum circuit  whose depth grows linearly with the system size (see Supplementary Section I.F)
\cite{Satzinger2021Realizing}:
\begin{equation}
    |E^{(0)}_+\rangle=\frac{1}{2^{4}}({1}+X_L)\prod_q({1}+B_q)|0\rangle^{\otimes18}.
\end{equation} 
We then measure the plaquette operators \{$A_p$\} and \{$B_q$\} (left panel of Fig.~\ref{fig3}{a}), and an average value of $\sim$0.95 is observed, which is noteworthy given that these operators encode four-body correlations. The high-fidelity gates and long coherence times achieved in our experiment are of crucial importance to obtain such a high average value of the measured stabilizers (see Supplementary Section II.B).  We also measure the expectation values of string operators $\{X_{L_i}, Z_{L_i}\}$ and find that $\langle E^{(0)}_+|Z_{L_i}|E^{(0)}_+\rangle\approx 0$ and $\langle E^{(0)}_+|X_{L_i}|E^{(0)}_+\rangle\approx 1$~(right panel of Fig.~\ref{fig3}{a}). 
These experimental results are in good agreement with theoretical predictions,  providing strong evidence that the prepared state is indeed a Floquet eigenstate as desired. %for the correctness of the prepared  Floquet eigenstate.

Having prepared the Floquet eigenstate, we further measure its topological entanglement entropy for two different subsystem sizes: four qubits and six qubits. We follow a protocol developed in Ref. \cite{Satzinger2021Realizing} and divide the subsystem into three parts: A, B, and C (upper panels of Fig.~\ref{fig3}{b}). $S_{\rm topo}$ can be extracted from the following combination of von Neumann entanglement entropies~\cite{Kitaev2006Topological, Levin2006Detecting}:
\begin{equation}
    S_\text{topo}=S_A+S_B+S_C-S_{AB}-S_{AC}-S_{BC}+S_{ABC},
\label{eq:S_topo}
\end{equation}
where $S_A$ is the von Neumann entropy for region $A$, while $AB$ means the union of regions $A$ and $B$, and similarly for other terms. For the eigenstates of $U_F$, the theoretically predicted value of $S_{\rm topo}$ is $-\ln{2}$~\cite{Kitaev2006Topological}. For each region $i$, we perform quantum state tomography on the whole (four-qubit or six-qubit) subsystem and reconstruct $\rho_i$ to calculate the corresponding fidelity $F(\rho_i)={\rm tr} \sqrt{\sqrt{\rho_i}\rho^{\rm ideal}_i\sqrt{\rho_i}}$ and von Neumann entropy $S_i=-{\rm tr}(\rho_i \ln \rho_i)$, where $\rho_i^{\rm ideal}$ is the reduced density matrix of region $i$ obtained by tracing out the complementary region of the ideal Floquet eigenstate.
The experimentally measured $S_{\rm topo}$,  $S_i$, and $F(\rho_i)$ are shown in the lower panels of Fig.~\ref{fig3}{b}.  The measured von Neumann entropy for each region agrees well with the corresponding ideal value. In addition, 
we observe that $-S_{\rm topo}/\ln2$ is  $0.86 \pm 0.02$ for the four-qubit and $ 0.84 \pm 0.08$ for the six-qubit subsystem, which is incompatible with the trivial-state value of zero and provides strong evidence for the nontrivial topological nature of the prepared Floquet eigenstate.  The deviation between the measured $S_{\rm topo}$ and its corresponding ideal value is due to limited coherence times and gate errors, which is confirmed by numerical results using a noise model estimated via independent measurements (our numerical simulations show that $-S_{\rm topo}/\ln2$ is $0.85$ and $0.82$ for the four-qubit and six-qubit subsystems, respectively; see Supplementary Section III).

\vspace{.5cm}
\noindent\textbf{\large{}Robustness against local perturbations}
{\large\par}
\noindent Topological order is expected to be robust against small local perturbations. In our experiment, we investigate the robustness of the subharmonic response of nonlocal logical operators and of the entanglement dynamics to local perturbations by turning on the random on-site fields in $H_1$. We vary the perturbation strength $B$ and measure  $\overline{A^{1/d}_{L_i}}(t)$ and $S_{\text{topo}}(t)$, with results plotted in Fig. \ref{fig4}.    

Figure \ref{fig4}{a} shows the measured disorder-averaged auto-correlation function $\overline{A^{1/d}_{L_i}}(t)$  for  $\{Z_{L_i}\}$ under weak~($B=0.1$) and strong~($B=3.0$) perturbations, which are averaged over 24 realizations with randomly drawn initial states,  $\alpha_p$, $\beta_q$, and ${\boldsymbol{B}_k}$. With a small perturbation ($B=0.1$), $\overline{A^{1/d}_{L_i}}(t)$ continues to exhibit persistent subharmonic response up to $20$ driving periods (upper panel of Fig.~\ref{fig4}{a}), which is a defining feature of the time-translation symmetry breaking for nonlocal operators and shows the robustness of the observed prethermal topologically ordered discrete time crystal. 
In contrast, with a strong perturbation ($B=3.0$), the measured $\overline{A^{1/d}_{L_i}}(t)$ decays quickly to zero and shows no subharmonic response~(lower panel of Fig.~\ref{fig4}{a}); at large $B$, the large onsite field rapidly destroys the topological order preventing any robust period doubling dynamics.
To explore the crossover from the time-crystalline to trivial dynamics, we vary the perturbation strength $B$ and Fourier transform the measured time-domain signals. % \lxiang{. Each time-domain signal for Fourier analysis is the average over $\{Z_{L_i}\}$}. 
Fig.~\ref{fig4}{b} shows the Fourier amplitudes at $\omega/\omega_0=0.5$ with $B$ ranging from $0$ to $3.0$. % \del{, averaged over  $\{Z_{L_i}\}$}. 
We find a small plateau at $B\lesssim0.25$, which further supports the robustness of the topologically ordered time-crystalline dynamics against weak perturbations.
As $B$ increases, the Fourier amplitude decays monotonically and becomes almost flat at $B\gtrsim2.5$, where the topological order is very quickly destroyed and no period doubling dynamics survives. % is observed. 
Sample-to-sample amplitude fluctuations over 24 random realizations (inset of Fig.~\ref{fig4}{b}) display a sharp increase at the same value of $B$ where the Fourier amplitude starts to decay, further highlighting the location of the crossover between the prethermal topological time-crystalline and the trivial dynamics.

We further study the dynamics of the topological entanglement entropy under local perturbations.  We first prepare the system in the $B=0$ Floquet eigenstate $|E^{(0)}_+\rangle$ and then let it evolve under $H(t)$ with varying $B$. In Fig.~\ref{fig4}{c}, we plot the measured  $\overline{S}_{\rm topo}(t)$ for $B=0.1$ and $B=3.0$, respectively.   
From this figure, we see that $\overline{S}_{\rm topo}(t)$ drops more quickly at strong perturbation $B=3.0$ due to the breakdown of the topological phase. We note that $\overline{S}_{\rm topo}(t)$ also has a slow decay even for $B=0.1$ due to the accumulated gate errors in the circuit, which is confirmed by the numerical simulations (the dashed lines in Fig.~\ref{fig4}{c})~(see Supplementary Section III). In addition, we measure plaquette and string operators after evolution under $U_F$ for a single time step (see Fig.~\ref{fig4}{d}). From this figure, it is clear that their values are largely preserved for $B=0.1$, unlike the case of $B=3.0$, where these values drop to near zero. % clearly different from that for $B=3$, where most initial information is lost. 
We further measure the disorder-averaged $\overline{S}_{\rm topo}(t=T)$ as a function of $B$ (Fig.~\ref{fig4}{e}). Similar to  the Fourier spectrum amplitudes in Fig.~\ref{fig4}{b},  $\overline{S}_{\rm topo}(t=T)$ also decays monotonically with increasing disorder strength. The plateau at weak disorder ($B\lesssim0.25$) further validates the robustness of time-crystalline dynamics against perturbations. % topological order against perturbations. 

We note that, for the generic local perturbations considered in our experiment, the overlap between a bare logical operator and its corresponding dressed logical operator may vanish in the thermodynamic limit. This would render the observation of time-crystalline behavior for the bare logical operator infeasible \cite{Wahl2021arxivTTC}.  In addition,
to observe the time-crystalline behavior, it is also crucial that $\{\prod \sigma^x_k, Z_L\}=0$ is satisfied, which requires that the length of $Z_L$ be odd. A possible way to maintain time-crystalline signatures in bare logical operators in the thermodynamic limit and to remove the requirement of odd length $Z_L$ is to consider a surface code with a hole, as discussed in depth in Ref.~\cite{Wahl2021arxivTTC}. In our experiment, we do not adopt such a layout because measuring  the corresponding nonlocal logical operator would become very challenging with the current device.%state-of-the-art superconducting qubit technologies. % in current stage. 

\vspace{.5cm}
\noindent \textbf{\large{}Conclusions and outlook}{\large\par}

\noindent In summary, we have experimentally observed signatures of a long-lived topologically ordered time crystal in the prethermal regime with a programmable superconducting quantum processor. In contrast to previously reported conventional time crystals, for our observed topologically ordered time crystal, the breaking of discrete time-translation symmetry only occurs for nonlocal logical operators, rather than local observables. We showed persistent subharmonic response for logical operators independent of the initial state and demonstrated  robustness of this response to generic perturbations without any microscopic symmetry. In addition, we also prepared a topologically ordered Floquet eigenstate and measured its topological entanglement entropy, which agrees well with theoretical predictions and clearly shows the intrinsic topological nature of the observed time crystal.

The topologically ordered eigenstates of the Floquet unitary are theoretically predicted to exhibit a perimeter law, where the expectation value of a Wilson loop scales with the perimeter rather than the area enclosed \cite{Wahl2021arxivTTC}. In the future, it is desirable to demonstrate such a perimeter law in experiment. The high controllability and programmability of the superconducting processor demonstrated in our experiment also paves the way to exploring a wide range of other exotic non-equilibrium phases with intrinsic topological order that are not accessible in natural materials. In particular, it would be interesting and important to realize various dynamically-enriched topological orders \cite{Potter2017Dynamically}. Indeed, our experiment has demonstrated all necessary building blocks for implementing the Floquet-enriched topological order that hosts dynamical anyon permutation \cite{Potter2017Dynamically} and emergent non-Abelian anyons~\cite{Wootton2008Non,Kalinowski2023NonAbelian}. An observation of such an unconventional phenomenon would also mark an important step in deepening our understanding of exotic non-equilibrium phases. 

\vspace{.5cm}
\noindent\textbf{\large Data availability}
\\The data presented in the figures and that support the other
findings of this study will be publicly available at Zenodo.org.

\vspace{.5cm}
\noindent\textbf{\large Code availability}
\\All the relevant source codes are available from the corresponding authors upon reasonable request.

%will be uploaded to some open repositories and will be publicly available upon the publication of the paper. 

\vspace{.5cm}
\noindent\textbf{Acknowledgement} The device was fabricated at the Micro-Nano Fabrication Center of Zhejiang University. We acknowledge the support from the Innovation Program for Quantum Science and Technology (Grant Nos. 2021ZD0300200 and 2021ZD0302203), the National Natural Science Foundation of China (Grant Nos.  92065204, 12274368, 12075128, T2225008), and Zhejiang Pioneer (Jianbing) Project (No. 2023C01036). T.I. acknowledges support from the National Science Foundation under Grant No.~DMR-2143635. F.M. acknowledges support from the NSF through a
grant for ITAMP at Harvard University. A.V.G.\ was supported in part by the NSF QLCI program (award No.~OMA-2120757). N.Y.Y. acknowledges support from the U.S. Department of Energy via the
National Quantum Information Science Research Centers Quantum Systems Accelerator and from a Simons Investigator award. 
W.J., S.J., W.L., Z.L., Z.-Z.S. and  D.-L.D. acknowledge in addition support from the Tsinghua University Dushi Program and Shanghai Qi Zhi Institute.

\vspace{.3cm}
\noindent\textbf{Author contributions}  L.X. and Z.B. carried out the experiments  and analyzed the experimental data under the supervision of Q.G. and H.W.;  W.J., L.X., S.J. and Z.B. performed the numerical simulations under the supervision of D.-L.D., Q.G., F.L., F.M., Z.-X.G., A.V.G., and T.I.; D.-L.D., W.J., F.L., F.M., Z.-X.G., A.V.G., N.Y., and T.I. conducted the theoretical analysis; H.L. and J.C. fabricated the device supervised by H.W.; D.-L.D., Q.G., W.J., L.X., H.W., F.L.,  Z.-X.G., A.V.G., F.M., and T.I. co-wrote the manuscript. 
All authors contributed to the discussions of the results.

\vspace{.3cm}
\noindent\textbf{Competing interests}  All authors declare no competing interests.

\bibliography{main_ref,Dengbib}

\end{document}

\nolinenumbers
\renewcommand{\bibnumfmt}[1]{[S#1]}
\renewcommand{\citenumfont}[1]{S#1}
 \setcounter{equation}{0}
    \setcounter{figure}{0}
    \setcounter{table}{0}
    
    \renewcommand{\theequation}{S\arabic{equation}}
    \renewcommand{\thefigure}{S\arabic{figure}}
\title{Supplementary Information: Long-lived topological time-crystalline order on a quantum processor}
\setcounter{figure}{0}
\setcounter{table}{0}
\renewcommand\thefigure{S\arabic{figure}}
\renewcommand\thetable{S\arabic{table}}

\maketitle
\tableofcontents

\section{Theoretical analysis}
\label{secs:theory}
In this work, we experimentally observe topological time-crystalline order, which can be characterized by the subharmonic temporal response of nonlocal logical operators \cite{Wen2017Colloquium, Wahl2021Topologically}.
%This is a new exotic many-body phenomenon in non-equilibrium systems.
The topological time-crystalline order is realized in a periodically driven surface-code model. %The intrinsic topological order of the surface-code model is inherited by the driving system. 
In this section, we will first briefly introduce the topological properties of the surface code and provide theoretical analysis of the emergence of topological time-crystalline order.  Then, we will show how to use a set of elementary quantum gates to implement the Floquet unitary and to prepare the Floquet eigenstates with a programmable superconducting quantum processor.  
% We also provide details of the quantum circuits.

\subsection{Surface-code model}
\begin{figure*}[tb]
    \includegraphics[width=0.8\textwidth]{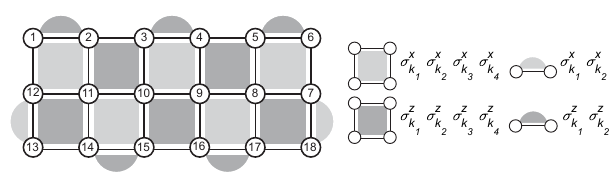}
\caption{\textbf{The layout of a  $3\times6$ rotated surface-code model.} Circles represent qubits. %, which are zig-zag indexed. 
The shaded plaquettes and semicircles indicate the local four-body and two-body operators on connected qubits, respectively. The dark (light) gray regions represent the $A_p$ ($B_q$) plaquette operators.
    \label{figs:layout}}
\end{figure*}
Recent progress \cite{Qi2011Topological,Senthil2015SymmetryProtected,Wen2017Colloquium} has demonstrated that many physical systems and their properties can be described using the language of topology; an important example of this is topological stabilizer codes \cite{Fujii2015Topological}. 
The surface-code model is an important topological stabilizer code that has numerous applications in quantum error correction \cite{Fowler2012Surface, Krinner2022Realizing, Acharya2022SuppressingQE,Bluvstein2023Logical}. It is analytically solvable and is also of interest to researchers from other fields, including condensed matter physics. 
%concerned by the broad researchers from many other communities such as condensed matter physics.

We adapt a variant of the surface-code model: the rotated surface code defined on a planar lattice with open boundary conditions \cite{Horsman2012Surface} (see Fig.~\ref{figs:layout}). Its Hamiltonian is given by 
\begin{equation}\label{eqs:hamiltonian}
    H=-\sum_p\alpha_p A_p-\sum_q\beta_q B_q,\ \text{with}\ A_p=\prod_{k\in p}\sigma_k^z\ \text{and}\ B_q=\prod_{k\in q}\sigma_k^x,
\end{equation}
where the plaquettes $p$ and $q$ are shown in  Fig.~\ref{figs:layout}, and $\alpha_p,\ \beta_q$ are randomly chosen positive coefficients.  For simplicity,  unless otherwise specified, we do not distinguish the plaquette operators and the semicircle operators in the following discussions.  We note that $A_p,\ B_q$ are both Pauli strings and have eigenvalues $\pm1$.  String operators consisting of the same type of Pauli operators commute with each other: $[A_p, A_{p'}]=[B_q, B_{q'}]=0$. Additionally, $A_p$ and $B_q$ also mutually commute because the overlap between the support of $A_p$ and any $B_q$ has an even number of qubits.
We conclude that the ground states of the rotated surface-code model are simultaneous eigenstates of all plaquette operators $A_p$ and $B_q$ with eigenvalue $+1$. 
Furthermore, each excited eigenstate of the Hamiltonian is a simultaneous eigenstate of all plaquette operators with different eigenvalues.
The model in Fig.~\ref{figs:layout} has 18 physical qubits and 17 independent plaquette operators, and the remaining single-qubit degree of freedom leads to a two-fold degeneracy for each energy level.

Suppose $|\psi\rangle$ is one of the ground states of $H$ defined on the lattice shown in Fig.~\ref{figs:layout}. That is, all plaquette operators have an expectation value of $1$: $\langle\psi|A_p|\psi\rangle=\langle\psi|B_q|\psi\rangle=+1$ for arbitrary plaquettes $p,q$.
To find another ground state, we try to flip the qubit with index $k=1$. However, this changes the sign of the expectation value of at least one of the plaquette operators, such as the one supported by qubits 1 and 2, i.e. $\langle\psi|\sigma^x_1A_{\{1,2\}}\sigma^x_1|\psi\rangle=-1$. Thus, the state with one flipped spin is no longer a ground state. 
To maintain the sign of $\langle A_{\{1,2\}}\rangle$, we flip qubit $2$, but this changes the signs of other connected plaquette operators. Thus we are forced to continue this process until qubits $k=1,\dots,6$ are all flipped. The final state is then still a simultaneous $+1$ eigenstate of all plaquette operators, since the string flip operator commutes with all plaquette operators: $\left[\prod_{k=1}^6\sigma_k^x, A_p\right]=\left[\prod_{k=1}^6\sigma_k^x, B_q\right]=0$. 
We denote the string operator by $X_L\equiv\prod_{k=1}^6\sigma_k^x$. It  cannot be represented as a product of any combination of plaquette operators. 
Similarly,  we can also define $Z_L\equiv\prod_{k=1,12,13}\sigma_k^z$, which anticommutes with $X_L$. 
This defines a Pauli algebra on a single logical qubit, and we conclude that the ground state manifold has a two-fold degeneracy indexed by, e.g., $\langle Z_L\rangle=\pm 1$.
The same analysis can also be applied to any excited eigenstate.  For the $l$-th energy level, we denote the eigenstates satisfying $\langle Z_L \rangle=\pm1$ by $|Z_L^{(l)}=\pm1\rangle$. In the topological stabilizer formalism,  plaquette operators are called stabilizers, and the ground space is called the code space, which is manipulated by logical string operators $X_L, Z_L$.

\subsection{Topological order of the surface-code model}
In the language of group theory, we can define a group generated by the plaquette operators (stabilizers) $S=\langle A_p, B_q\rangle$, and an 18-qubit Pauli group $P_{18}=\langle \sigma_k^x,\sigma_k^z\, : \text{$k$ runs over all sites}\rangle$. $S$ is a subgroup of $P_{18}$. The centralizer group $C_{P_{18}}(S)$ consists of all operators in $P_{18}$ that simultaneously commute with all plaquette operators. For the rotated surface-code model, $S$ is a normal subgroup of $C_{P_{18}}(S)$ and we have the quotient group $C_{P_{18}}(S)/S\cong P_1$ being the Pauli group for one qubit. 
%So, the eigenspace is two-fold degenerate. 
The above discussion indicates that the two-fold degeneracy of each eigenstate of Eq.~\eqref{eqs:hamiltonian}, discussed in the previous section, is directly related to the topology of the system. 
Any single-qubit operator fails to commute with at least one plaquette operator.  According to the discussion in Section \ref{secs:theory}{.A}, we know that $Z_L$ and $X_L$ are two independent nonlocal operators which can map one eigenstate to its degenerate partner. As a consequence, they can be regarded as the representatives of the cosets of $S$. Since arbitrary products of the plaquette operators with $Z_L$ ($X_L$) are equivalent to $Z_L$ ($X_L$) itself in a fixed stabilizer eigenspace, we have several equivalent expressions for operators  $Z_L$ and $X_L$ (see Fig.~2  of the main text):
$Z_{L_1}=\prod_{k=1,12,13}\sigma_k^z,\ Z_{L_2}=\prod_{k=2,11,14}\sigma_k^z, \dots$ and $X_{L_1}=\prod_{k=1}^6\sigma_k^x,\ X_{L_2}=\prod_{k=7}^{12}\sigma_k^x,\dots$, all of which are nonlocal operators.  %It is equivalent to choose any one of the operators from each coset.
All operators within each coset are equivalent modulo $S$.

A connection between topological order and quantum entanglement is provided by the topological entanglement entropy.
For a many-body wavefunction,  the von Neumann entropy of a subregion, $S(\rho_\text{sub}) \equiv-\operatorname{tr} \rho_\text{sub} \ln \rho_\text{sub}$, describes the quantum entanglement between the subregion and its complement. Here, $\rho_\text{sub}$ denotes the reduced density matrix of the subregion obtained by tracing out the complementary region.
\begin{figure}[tb]
    \includegraphics[width=0.4\textwidth]{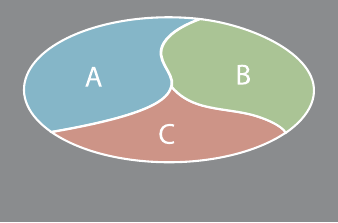}
    \caption{Regions $A$, $B$, and $C$ defined for computing topological entanglement entropy via Eq.\ (\ref{eqs:topo_entropy_computing}). }
    \label{figs:regions}
\end{figure}
For a system obeying the entanglement area law, the von Neumann entropy of a subregion is represented as \cite{Kitaev2006Topological}
\begin{equation}\label{eqs:topo_entropy}
    S(\rho_\text{sub})=\alpha \partial_\text{sub} -\gamma+\cdots.
\end{equation}
Here, $\alpha$ is a constant determined by the details of the system, $\partial_\text{sub}$ is the volume of the subregion's boundary, and $-\gamma$ is the topological entanglement entropy, which characterizes nonlocal entanglement persisting at arbitrarily large distances. 
In topological quantum field theory, it is known that $\gamma=\ln \mathcal{D}$, where $\mathcal{D}$ is the total quantum dimension~\cite{Kitaev2006Topological}. In an Abelian anyon model, $\mathcal{D}$ is the square root of the number of superselection sectors corresponding to inequivalent quasi-particle species. For the rotated surface-code model, there are two types of Abelian anyons: electric charges $e$ associated with $A_p$ plaquette operators and magnetic charges $m$ associated with $B_q$ plaquette operators. So, there are four quasi-particle sectors (identity, $e$, $m$, and $em$) and the total quantum dimension $\mathcal{D}$ is equal to $\sqrt{4}=2$. Thus, the topological entanglement entropy $S_{\text{topo}}=-\ln2$ for the model~\eqref{eqs:hamiltonian}.
To measure $S_{\text{topo}}$, we cancel out boundary contributions by dividing the subregion into three parts that are all large compared to the correlation length \cite{Kitaev2006Topological} (see Fig.~\ref{figs:regions}) and computing
\begin{equation}\label{eqs:topo_entropy_computing}
    S_{\text {topo }} = S_A+S_B+S_C-S_{A B}-S_{B C}-S_{A C}+S_{A B C},
\end{equation}
where $S_A$ is the von Neumann entropy of region $A$, $S_{AB}$ is the von Neumann entropy of region $A\cup B$, and so on.  In this way, all boundary terms are canceled out, and the result is the topological entanglement entropy $S_{\text {topo }}=-\gamma$.

\subsection{Topological time-crystalline order}
\label{sec:topo_floquet_system}
\begin{figure*}
    \centering
    \includegraphics[width=\textwidth]{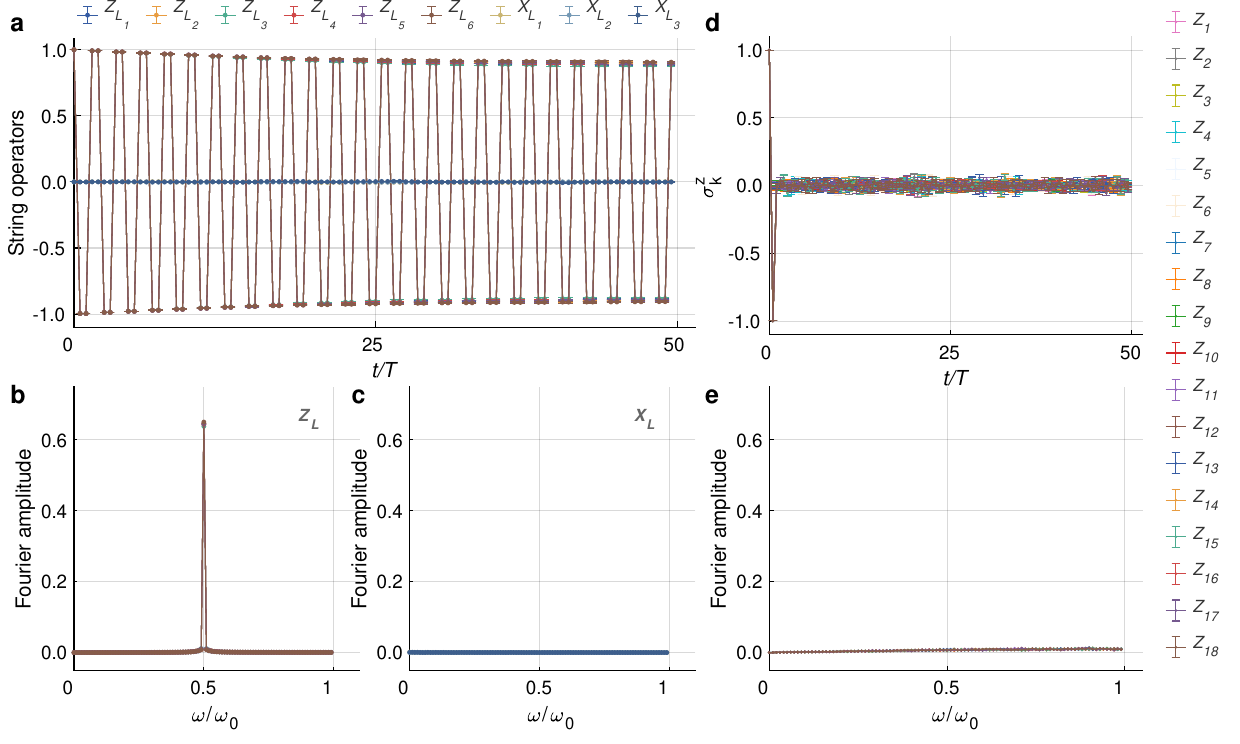}
    \caption{\textbf{Dynamics of nonlocal and local observables at $B=0.1$.} All data in this figure are averaged over 1000 random realizations, and the error bars stand for the standard error of the statistical mean.
    {\bf a}, Disorder-averaged dynamics of nonlocal string operators $\{Z_{L_i}\}$ and $\{X_{L_i}\}$.  Expectation values of  all six $\{Z_{L_i}\}$ lying on top of each other, break the time-translation symmetry, and manifest subharmonic oscillations with period $2T$. Despite the initial slight decay, mainly caused by the imperfect overlap between the dressed logical operators and the measured bare operators, the $\{Z_{L_i}\}$ expectation values show a plateau at late times, indicating persistent oscillations. In contrast, expectation values of all three $\{X_{L_i}\}$ (lie on top of each other) remain zero as expected.
    {\bf b}, The Fourier spectrum of the disorder-averaged dynamics of $\{Z_{L_i}\}$. The peaks at $\omega/\omega_0=1/2$ ($\omega_0=2\pi/T$) indicate the subharmonic oscillations of $\{Z_{L_i}\}$.
    {\bf c}, The Fourier spectrum of the disorder-averaged dynamics of $\{X_{L_i}\}$. 
    {\bf d}, Disorder-averaged dynamics of single-qubit operators $\{\sigma^z_k\}$. 
    {\bf e}, The Fourier spectrum of the disorder-averaged dynamics of single-qubit operators $\{\sigma^z_k\}$.}
    \label{figs:dynamics}
\end{figure*}

Having reviewed the concept of topological order in the static rotated surface-code model, we now generalize to the periodically driven setting and show the emergence of topological time-crystalline order. The Floquet Hamiltonian of the driven rotated surface-code model is
\begin{equation}\label{eqs:floquet_model}
    \begin{aligned}
    H(t) & = \begin{cases}H_1, & 0 \leq t<T^{\prime}, \\
    H_2, & T^{\prime} \leq t<T,\end{cases} \\
    H_1 & \equiv\frac{\pi}{2} \sum_k\sigma_k^x+\sum_k\boldsymbol{B}_k\cdot\boldsymbol{\sigma}_k, \\
    H_2 & \equiv-\sum_p\alpha_pA_p-\sum_q\beta_qB_q,
    \end{aligned}
\end{equation}
where $\boldsymbol{B}_k$ is an on-site field randomly chosen from a ball with radius $B$, $\boldsymbol{\sigma}_k$ is the vector of Pauli operators, $A_p, B_q$ are plaquette operators defined in Fig.~\ref{figs:layout}, $\alpha_p, \beta_q$ are coefficients uniformly chosen from $[0, 2\pi)$, and $T=2T'=2$.
$H_2$ is the rotated surface-code Hamiltonian, and its spectrum is exactly two-fold degenerate, with eigenstates $|Z_L^{(l)}=\pm1\rangle$ at  energy $\epsilon_l$. These two-fold degenerate eigenstates cannot be distinguished by any local operator. However, they can be distinguished by a nonlocal string operator such as $Z_L$. When $B=0$, the effect of $U_1=e^{-i\pi/2\sum_k\sigma_k^x}\propto\prod_k\sigma_k^x$ is to flip all the qubits.
%(Here, we choose $B=0$ for simplicity). 
When we index degenerate eigenstates of $H_2$ with a logical operator $Z_L$ of odd length, this is equivalent to applying a logical operator $X_L$.
In this situation, we have the relations 
\begin{equation}
    U_F|Z_L^{(l)}=1\rangle=\exp{(-i\epsilon_l)}|Z_L^{(l)}=-1\rangle, \quad U_F|Z_L^{(l)}=-1\rangle=\exp{(-i\epsilon_l)}|Z_L^{(l)}=1\rangle,
\end{equation}
% $$U_F|Z_L^{(l)}=1\rangle=\exp{(-i\epsilon_l)}|Z_L^{(l)}=-1\rangle, \quad U_F|Z_L^{(l)}=-1\rangle=\exp{(-i\epsilon_l)}|Z_L^{(l)}=1\rangle,$$
where $U_F=\exp{\left(-iH_2\right)}\exp{\left(-iH_1\right)}$ is the Floquet unitary. In other words, $U_F$ toggles between two degenerate eigenstates of $H_2$. In a degenerate eigenspace of $H_2$, $U_F$ can thus be represented as a matrix
\begin{equation}
    U_F\sim
\left[\begin{array}{cc}
0 & \exp \left(-i \epsilon_l\right) \\
\exp \left(-i \epsilon_l\right) & 0
\end{array}\right].
\end{equation}

Therefore, in this eigenspace, $U_F$ has eigenvalues $\pm \exp \left(-i \epsilon_l\right)$ corresponding to Floquet eigenstates $|E_\pm^{(l)}\rangle\propto|Z_L^{(l)}=1\rangle\pm|Z_L^{(l)}=-1\rangle$, respectively. For the Floquet Hamiltonian $H_F=i\log U_F$, the corresponding quasi-energies are $\epsilon_l$ and $\epsilon_l+\pi$ (see Fig.~1b of the main text).
We note that the $H_2$ eigenstates $|Z_L^{(l)}=1\rangle$ and $|Z_L^{(l)}=-1\rangle$ have the same topological entanglement entropy $S_\text{topo}=-\ln{2}$, which is the core feature of the topological order of this model. The Floquet eigenstates $|E^{(l)}_\pm\rangle$ inherit the same value of $S_{\text{topo}}$.

We further investigate the dynamical behavior of the nonlocal string operators.  Without loss of generality, we start from a product state $\left|\psi_0\right\rangle$ which has expectation value of $+1$ for the string operator $Z_L$, such as $\ket{\psi_0}=\bigotimes_k\ket{0}_k$. This can be represented as a superposition of the subset of eigenstates $\{|Z_L^{(l)}=1\rangle\}$: $|\psi_0\rangle=\sum_l\alpha_l|Z_L^{(l)}=1\rangle$, such that$\langle\psi_0|Z_L|\psi_0\rangle=\sum_{l'l}\alpha^*_{l'}\alpha_{l}\langle Z_L^{(l^\prime)}=1|Z_L|Z_L^{(l)}=1\rangle=\sum_{l'l}\alpha^*_{l'}\alpha_{l}\delta_{l'l}=1$. Under time evolution by the Floquet unitary $U_F$, we find that 
\begin{equation}
\begin{split}
    U_F|\psi_0\rangle&=\exp{(-iH_2)}\exp{(-iH_1)}\sum_l\alpha_l|Z_L^{(l)}=1\rangle\\
    &=\exp{(-iH_2)}\sum_l\alpha_l|Z_L^{(l)}=-1\rangle\\
    &=\sum_l\alpha_le^{i\epsilon_l}|Z_L^{(l)}=-1\rangle\\
    &\equiv|\psi_1\rangle.
\end{split}
\end{equation}
If we measure the string operator $Z_L$ after a single Floquet period $T$, we have 
\begin{equation}
    \langle\psi_1|Z_L|\psi_1\rangle=\sum_{l'l}\alpha^*_{l'}\alpha_{l}e^{-i(\epsilon_l-\epsilon_{l'})}\langle Z_L^{(l^\prime)}=-1|Z_L|Z_L^{(l)}=-1\rangle=-\sum_{l'l}\alpha^*_{l'}\alpha_{l}e^{-i(\epsilon_l-\epsilon_{l'})}\delta_{l'l}=-1.
\end{equation}
Similarly, the stroboscopic dynamics of the string operator $Z_L$ after a time $t=nT$ is $\langle\psi_n|Z_L|\psi_n\rangle=(-1)^n$,  where $|\psi_n\rangle=\left(U_F\right)^n|\psi_0\rangle$. Therefore, the expectation values of $Z_L$ oscillate with period $2T$,  which breaks the discrete time-translation symmetry of the Floquet Hamiltonian (\ref{eqs:floquet_model}). The corresponding numerical simulations are shown in Fig.~\ref{figs:dynamics}{a}, {b}. For the string operator $X_L$, one can check that $\langle\psi_n|X_L|\psi_n\rangle=0$ (see Fig.~\ref{figs:dynamics}{a}, {c}).

In contrast, we find that local operators exhibit featureless dynamics in our Floquet model.  For a single-qubit operator $O_\text{single}$, we have 
\begin{equation}
    \langle\psi_0|O_\text{single}|\psi_0\rangle=\sum_{l'l}\alpha^*_{l'}\alpha_{l}\langle Z_L^{(l^\prime)}=1|O_\text{single}|Z_L^{(l)}=1\rangle,
\end{equation}
and 
\begin{equation}
    \langle\psi_1|O_\text{single}|\psi_1\rangle=\sum_{l'l}\alpha^*_{l'}\alpha_{l}e^{-i(\epsilon_l-\epsilon_{l'})}\langle Z_L^{(l^\prime)}=-1|O_\text{single}|Z_L^{(l)}=-1\rangle. 
\end{equation}
% Because no single-qubit operator can distinguish topologically ordered eigenstates  $|Z_L^{(l)}=1\rangle$ and $|Z_L^{(l)}=-1\rangle$, the extra phase factor $e^{-i(\epsilon_l-\epsilon_{l'})}$ for each component cannot be canceled out. Instead, t
The extra phase factors $e^{-i(\epsilon_l-\epsilon_{l'})}$ tend to be randomly distributed under the Floquet dynamics of Hamiltonian (\ref{eqs:floquet_model}), leading to a fast decay to zero for $\langle\psi_n|O_\text{single}|\psi_n\rangle$ (see Fig.~\ref{figs:dynamics}{d}, {e}) and thus to an impossibility of time-translation symmetry breaking. This distinction between the dynamics of local and nonlocal operators is the key difference between the topological time-crystalline order and conventional time-crystalline order.

\subsection{Stability of topological time-crystalline order}

\begin{figure*}[tb]

    \centering
    \includegraphics[width=0.9\textwidth]{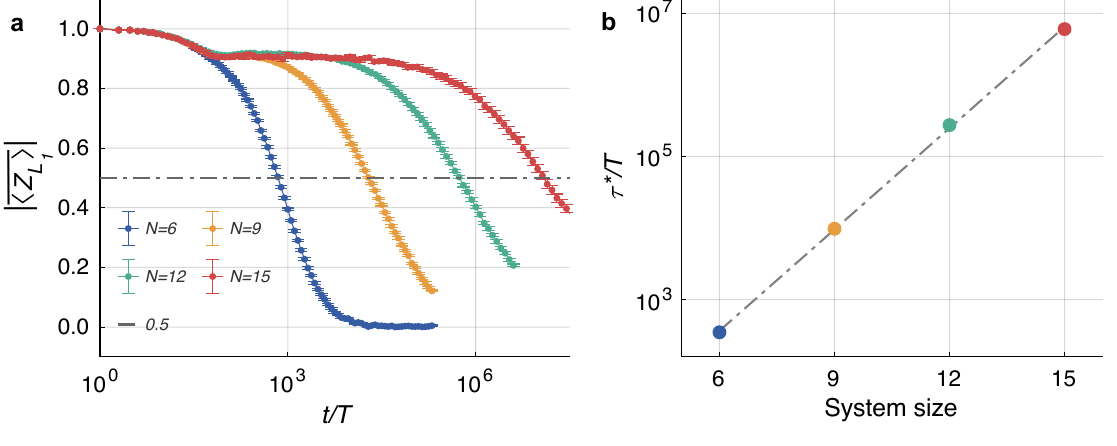}
    \caption{\textbf{Lifetime of the topological time-crystalline order.} Experimentally measuring the lifetime of the topological time-crystalline order is infeasible due to the limited coherence time of our experimental device.
    {\bf a}, Disorder-averaged dynamics of the nonlocal logical qubit expectation values $\Bigl\lvert\overline{\langle\psi_n|Z_{L_1}|\psi_n\rangle}\Bigl\rvert$ (we first take the average, and then take the absolute value).  The results are averaged over a number of random realizations ranging from  $10^3$  ($N=15$) to $10^4$ ($N=6$) depending on the system size. The results show that, after an initial slight decay, $\Bigl\lvert\overline{\langle\psi_n|Z_{L_1}|\psi_n\rangle}\Bigl\rvert$ reaches a plateau extending up to a timescale that diverges exponentially with the system size (see panel {\bf b}). The gray dashed line indicates the reference value of $1/2$ that is used to extract the logical-qubit lifetime.  Error bars represent the standard error of the statistical mean.
    {\bf b}, Finite-size scaling of the lifetime of the topological time-crystalline order. The colored dots exhibit the exponential scaling of $\tau^*/T$ with system size, where $\tau^*$ is the time at which $\Bigl\lvert\overline{\langle\psi_n|Z_{L_1}|\psi_n\rangle}\Bigl\rvert$ reaches $1/2$. The gray dashed line is a best-fit exponential for the system-size dependence of $\tau^*/T$.}
    \label{figs:lifetime}
\end{figure*}

The stability of the topological time-crystalline order is a highly non-trivial problem and is still open for further exploration. For example, it is believed that many-body localization (MBL) can help prevent a Floquet system from heating to infinite temperature. However, the stability of MBL in the thermodynamic limit is still controversial~\cite{Doggen2020Slow,Potirniche2019Exploration,DeRoeck2017Manybody} (especially in two or more spatial dimensions), and investigating such open problems is beyond the scope of the present work. Here, we perform a preliminary numerical simulation to explore the stability of the topological time-crystalline order against small local random fields.

In our numerical simulations,  we define the lifetime as the number of Floquet periods elapsed before the magnitude of the expectation value of $Z_L$ decays to $1/2$. We fix a small perturbation strength $B=0.1$ for this study. In the presence of random on-site fields, the contribution of $H_1$ to the Floquet unitary is no longer a perfect spin-flip operator for all the sites; rather, it is
\begin{equation}
    U_1=e^{-iH_1}\sim\prod_k\sigma_k^x+O(B).
\end{equation} 
%As before, we also add random disorder to $H_2$ by uniformly sampling the coefficients $\alpha_p, \beta_q$ in Eq.~\eqref{eqs:floquet_model} from $[0, 2\pi)$.
As the underlying lattice structure can have a significant effect on the lifetime of the logical qubit, we carry out the numerical simulations for lattices of dimensions $3\times2,3\times3,3\times4,3\times5$, and measure the non-local logical $Z$ operator of fixed length $3$.  The initial states are randomly chosen $z$-basis product states in order to give definite expectation values of $\pm 1$ for the string operator $Z_L$.  For each realization, we numerically calculate the dynamics under Eq.~(\ref{eqs:floquet_model}) for a long enough time to observe the decay of the string operator.  The final results are averaged over many such random realizations, and the time at which the expectation values reach $1/2$ is regarded as the indicator of the lifetime (see Fig.~\ref{figs:lifetime}{a}).  Our results show that the lifetime of the topological order for the model is much longer than any experimentally accessible timescale. Furthermore, we observe that the lifetime of the logical qubit grows exponentially with system size for $N=6,9,12,15$~(see Fig.~\ref{figs:lifetime}{b}). We conclude that, owing to disorder, the true thermalization time is much larger than these experimentally inaccessible timescales, and thus the observed topological time crystalline behavior is robust.

\subsection{Quantum circuits for the Floquet unitary}
\label{sec:circuit_of_time_evo}
\begin{figure*}[t]
    \centering
    \includegraphics{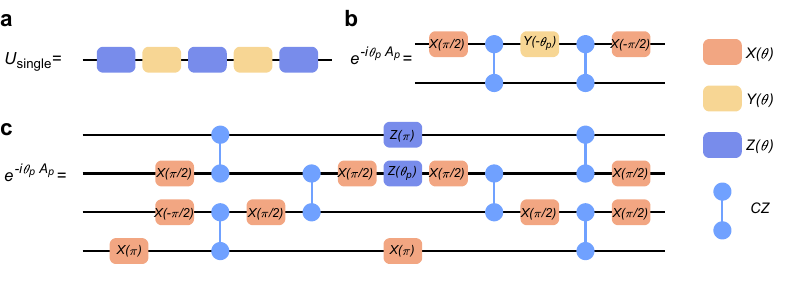}
    \caption{\textbf{Digital quantum circuits for the Floquet unitary.} 
    ({\bf a}) Single-qubit circuit realizing $U_1$. We use Euler angles to represent  general single-qubit rotations.
    ({\bf b}) Circuit realizing the evolution unitary of the two-body plaquette $A_p$ on the boundary of the lattice (see Fig.~\ref{figs:layout}).
    ({\bf c})  Circuit realizing the evolution unitary of the four-body plaquette $A_p$. Combining these elementary circuits allows us to digitally simulate the Hamiltonian (\ref{eqs:floquet_model}). In experiments, the whole circuit is further compiled to reduce the depth.}
    \label{figs:circuits}
\end{figure*}

It is straightforward to realize the Floquet drive  $U_1(t)=e^{-itH_1}$ with tensor products of single-qubit rotations, which can be represented via Euler angles (see Fig.~\ref{figs:circuits}{a}). However,  the circuit construction of $U_2(t)=e^{-itH_2}$ is more challenging due to the four-body plaquette operators $A_p$, $B_q$ in the Hamiltonian $H_2$. To construct digital quantum circuits for this evolution, we exploit the property that all plaquette operators mutually commute, so that the evolution reads
\begin{equation}
    U_2(t)=e^{-itH_2}=\prod_pe^{it\alpha_pA_p}\prod_qe^{it\beta_qB_q}.
\end{equation}

Besides, we have the relation $H\sigma^zH=\sigma^x$, where $H$ is the single-qubit Hadamard gate. Thus, for any plaquette $q$, we have 
\begin{equation}
    H^{\otimes q}\left(e^{it\beta_qA_q}\right)H^{\otimes q}=e^{it\beta_qB_q},
\end{equation} 
where $H^{\otimes q}$ stands for a Hadamard transform applied to the qubits in plaquette $q$.
Therefore, the circuit construction for evolution under $H_2$ reduces to simulating a single plaquette operator $A_p$. 

Variational quantum circuits are a powerful tool for NISQ quantum computation and quantum simulation and have been intensively studied in recent years \cite{Cerezo2021Variationalb, Li2022Quantuma}. We adapt this method to construct the quantum circuit for evolution under the plaquette operator $A_p$.  Variational quantum circuits are composed of gates with parameterized rotation angles that can be updated according to various algorithms. %For more details on variational quantum circuits, we refer the reader to Refs.~. 
The circuit construction for the evolution operator of $A_p$ can be divided into two steps. First, we need to find an appropriate variational ansatz for it. Second, we optimize the variational parameters in this ansatz to minimize the distance between the corresponding quantum circuit and the target unitary.

For the first step, we use the neuroevolution method \cite{Lu2021Markovian} to find a suitable variational quantum circuit architecture. The complete gate set used in our experiment consists of three kinds of single-qubit rotations $X(\theta), Y(\theta), Z(\theta)$ and a controlled-phase gate ${\rm CR}_z(\theta)$ along the $z$ axis ($\theta$ stands for the variational parameter). Then, we can construct a directed graph where each node represents a block of gates that can be implemented in parallel, and where the directed edges denote allowed sequences of blocks. A quantum circuit can then be represented as a directed path in this graph. 
To find a desired circuit ansatz, we use the following procedure: 
\begin{enumerate}
    \item Randomly sample several paths with a fixed depth in the constructed directed graph as our initial quantum-circuit ansatz;
    \item For each path representing an ansatz, minimize the cost function $L(\boldsymbol{\theta})$, which is given by  
    \begin{equation}
        L(\boldsymbol{\theta})=1-\operatorname{Tr}\left[U_{\text {target }}^{\dagger} U_{\text {circuit }}(\boldsymbol{\theta})\right] / d,
    \end{equation}
    where $U_{\text {target }}$ is the evolution unitary of the plaquette operator $A_p$,  $U_{\text {circuit }}$ is the unitary represented by the current quantum-circuit ansatz with variational parameters $\boldsymbol{\theta}$, and $d$ is the dimension of the  Hilbert space. $L(\boldsymbol{\theta})$ measures the distance between the target unitary and the current quantum circuit. We update variational parameters using gradient-based algorithms; 
    \item Keep the ansatzs with smallest values of the loss function $L(\boldsymbol{\theta})$, and prolong the corresponding paths in the directed graph to generate new circuits with larger depth; 
    \item Iterate steps 2 and 3 until the loss function converges.
\end{enumerate}
We repeat this algorithm several times and choose the circuit ansatz with the smallest value of the loss function as the optimal result.  

Remarkably, in some cases, we can obtain ansatz circuits with an extremely small distance from our target unitary (typically, the loss function $L(\boldsymbol{\theta})<10^{-4}$), which indicates that there probably exists a circuit ansatz that is analytically equivalent to the target unitary. Thus, we further manually simplify the obtained variational circuit ansatz. Precisely speaking, we alternately utilize the following methods to reduce the number of the variational parameters in this ansatz:
\begin{enumerate}
    \item Drop those gates very close to the identity gate, i.e., variational gates with small rotation angles.
    \item Fix those gates with special parameters, such as $\theta=\pi$. 
    \item Change the order of some commuting gates.
    \item Split or combine some neighboring gates.
\end{enumerate}
After the reduction process above, we obtain an experimentally friendly circuit that analytically represents the evolution unitary $U(t)=e^{it\alpha_pA_p}$ (see Fig.~\ref{figs:circuits}{b}, {c}). The evolution unitary of $B_q$ is then obtained by inserting a layer of Hadamard gates before and after $U(t)$.  Then,  we can obtain the digital quantum circuit for $U_2$ by concatenating the evolution operators for all plaquette operators. This yields an analytical representation of the whole time-evolution unitary (see Fig.~1 of the main text). 

\subsection{Floquet eigenstate preparation circuits}
\label{sec:circuit_of_eigenstate}
In the main text, we experimentally measure the topological entanglement entropy of a Floquet eigenstate. Here, we provide more details on the circuit for preparing this state. In our model, the Floquet eigenstates  $|E_\pm^{(l)}\rangle\propto |Z_L^{(l)}=1\rangle\pm|Z_L^{(l)}=-1\rangle$ are superpositions of the eigenstates ($|Z_L^{(l)}=\pm1\rangle$) of the rotated surface-code Hamiltonian.  We follow the method in Ref.~\cite{Satzinger2021Realizing}  to realize the eigenstates  $|Z_L^{(l)}=\pm1\rangle$ and their superpositions. Here, we briefly summarize the idea of this method.

Without loss of generality, we choose the Floquet eigenstate that superposes the two-fold degenerate ground states of $H_2$.  Because the ground states are the simultaneous eigenstates of all plaquette operators, we can apply the projector $\prod_{p,q}(\mathbbm{1}+A_p)(\mathbbm{1}+B_q)$ to map the initial state into the ground space.  For simplicity, we choose an initial state of $|0\rangle^{\otimes 18}$.  The projector above is a sequence of mutually commuting projection operators, which project the initial state into a simultaneous eigenstate of $A_p$ and $B_q$. The final state is unchanged after exchanging the order of the projection operators, since $[A_p, B_q]=0$.  Since $(\mathbbm{1}+A_p)|0\rangle^{\otimes 18}=|0\rangle^{\otimes 18}$, we conclude that $|Z_L^{(0)}=1\rangle\propto\prod_{q}(\mathbbm{1}+B_q)|0\rangle^{\otimes 18}$ (ignoring an overall normalization factor). Furthermore, 
%due to $|Z_L^{(0)}=1\rangle=X_L|Z_L^{(0)}=-1\rangle$ and 
since $|Z_L^{(0)}=-1\rangle=X_L|Z_L^{(0)}=1\rangle$, we can express the superposition of ground states in two topological sectors as $|Z_L^{(0)}=1\rangle+|Z_L^{(0)}=-1\rangle\propto(\mathbbm{1}+X_L)\prod_q(\mathbbm{1}+B_q)|0\rangle^{\otimes18}$. Since $B_q$ only consists of $\sigma^x$ operators, we have $[X_L,B_q]=0$.   As a result, we can also write $|Z_L^{(0)}=1\rangle+|Z_L^{(0)}=-1\rangle\propto\prod_q(\mathbbm{1}+B_q)(\mathbbm{1}+X_L)|0\rangle^{\otimes18}$. 

Since all the $X_{L_k}$ are equivalent in our model, we fix $X_L=\prod_{k=7}^{12}\sigma_k^x$ in the following discussion.
We first show how to prepare the state $(\mathbbm{1}+X_L)|0\rangle^{\otimes18}$.  The effect of $\mathbbm{1}+X_L$ is to project the initial zero state into a cat state of the form$(|000000\rangle+|111111\rangle)_{7,\dots,12}\otimes|0\rangle^{\otimes12}$.  This cat state can be prepared conveniently with quantum gates by first applying a Hadamard gate on one of the qubits in $\{7,\dots12\}$ and then successively applying CNOT gates for each pair of neighboring qubits. The effect of the projection operators $\mathbbm{1}+B_q$ can be realized by applying similar gate sequences to the qubits in plaquette $q$.

\section{Experimental information}
\label{secs:experiment}
Exploring  an intrinsically non-equilibrium Floquet system [Eq.~\eqref{eqs:floquet_model}] relies on the dynamical manipulation of highly entangled many-body states.
%, which is beyond the capacity of the known natural materials.
In our work, we engineer a topological ``synthetic quantum material" on a $3\times6$ superconducting qubit lattice using the digital quantum simulation paradigm.  In this section, we provide detailed information on the experimental platform, device performance, and circuit calibration.

\subsection{Experimental platform}
\label{secs:chip_characterization}
\begin{figure*}[h]
\center
\includegraphics[width=1.0\linewidth]{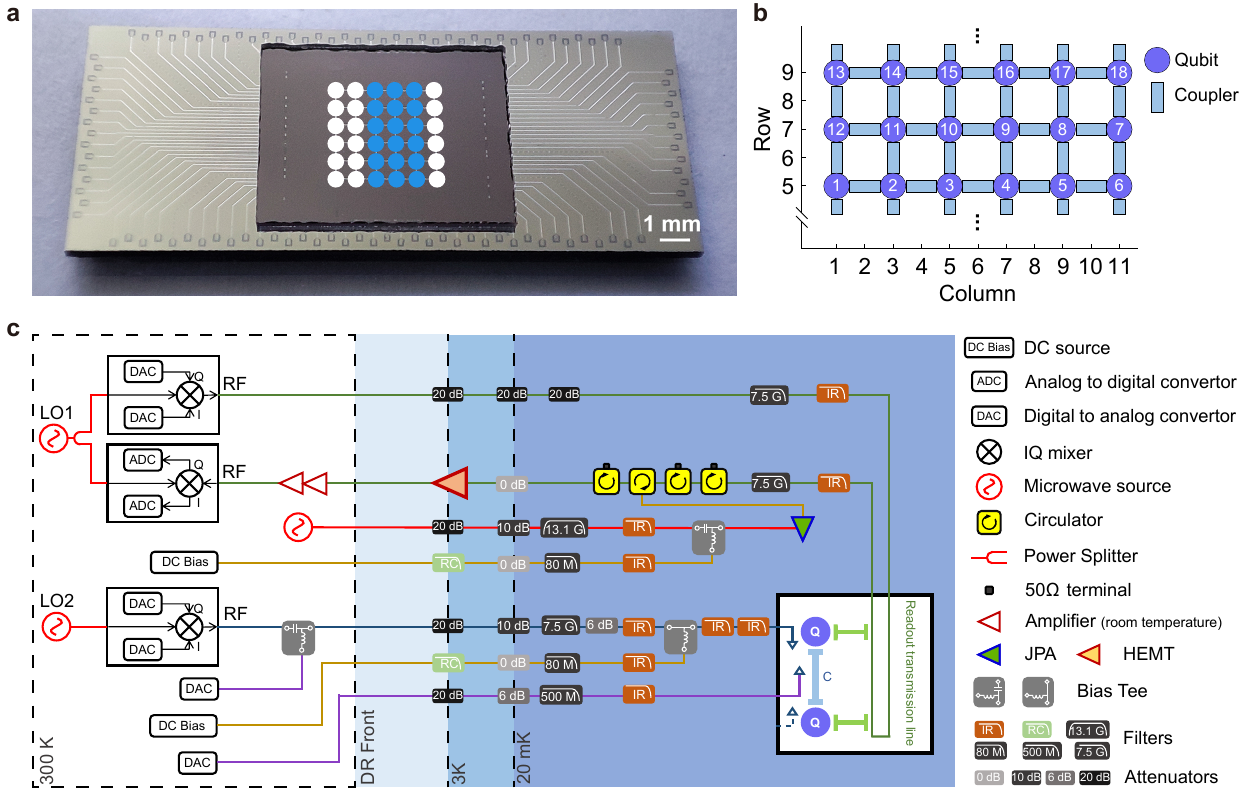}
\caption{{\bf Quantum processor and experimental setup.} 
{\bf a}, Photo of the flip-chip quantum processor. The 18 qubits actively used in our experiments are marked by blue solid circles, while the other unused qubits are marked by white circles.  
{\bf b}, Schematic structure of the $3\times6$ qubit lattice. Qubits are arranged at the vertices of a 3$\times$6 grid, and couplers are represented by the edges.
{\bf c}, Wiring information. The superconducting quantum chip is mounted on the mixing chamber plate ~(20 mK) of the dilution refrigerator.  For simplicity, we use a square box at the bottom right corner,  wherein a pair of qubits is coupled to a coupler, to represent the chip. To control and measure the chip, qubits and couplers are connected to the room-temperature electronics by readout lines (green), microwave-drive lines (blue), fast Z-pulse lines (purple) and slow DC-bias lines (brown).  Information on the microwave components is provided in the legend on the right.
}
\label{fig:chip_wiring}
\end{figure*}
As shown in Fig.~\ref{fig:chip_wiring}{a} and {b},  we use a $3\times6$ qubit lattice on a flip-chip superconducting quantum processor to implement the theoretical model.  This processor has a two-dimensional architecture consisting of a $6\times6$ qubit array where the nearest-neighbor (NN) qubit pairs are connected by tunable couplers~\cite{Yan2018PRApp}. Thus, the interaction between each NN qubit pair can be tuned dynamically by applying control signals to the coupler (Fig.~\ref{fig:chip_wiring}{c}), which enables the implementation of two-qubit CZ gates in our experiments. Each qubit is capacitively coupled to a readout resonator for dispersive readout, and each group of nine readout resonators shares a common readout transmission line for simultaneous state measurements. Figure~\ref{fig:chip_wiring}{c} shows the details of the experimental setup, including wiring information,  microwave components, and room-temperature control electronics. 

\subsection{System calibration}
\label{SystemCalibration}
Realizing a well-specified Hamiltonian with digital quantum circuits is a challenging experimental task. It is a complex control problem, whose target is to find optimal parameters which convert time-dependent room-temperature microwave signals to an effective Hamiltonian on the quantum chip at low temperature. In our experiment, we coherently control 45 quantum elements (18 qubits and 27 couplers) for manipulating the topologically ordered Floquet system. Here, we briefly describe our calibration procedures for tuning up these elements.  

Before calibrating the universal quantum gate set (single-qubit rotations and two-qubit CZ gate),  we use the following procedures to get an overall characterization of the device.  
\begin{enumerate}
\item Perform spectroscopy measurements for each qubit to obtain the relationship between the qubit frequency $\omega_{q}$ and the amplitude of its fast Z pulse. 
\item Tune up each qubit individually at a flux-sensitive point, which we choose $\sim$300 MHz below its maximum frequency (sweet point) in our experiments, and perform a series of measurements to obtain the following basic parameters.
    \begin{itemize}
    \item Single-qubit $\pi$ and $\pi/2$ pulse parameters.
    \item The ratio between the fast Z-pulse amplitude and the slow DC-bias amplitude. With this ratio, we can further get the relationship between qubit frequency $\omega_{q}$ and the amplitude of the DC bias.
    \item Qubit readout pulse parameters.
    \item Spectrum of $T_1$. We note that the $T_1$ spectrum is repeatedly monitored on different days to detect possible moving two-level-system (TLS) defects~\cite{Klimov2018}.
    \end{itemize}

\item Synchronize the timing of the control pulses from different control lines.  We select the center qubit as the root and use the Dijkstra algorithm to traverse all the qubits and couplers from near to far.  For the detailed calibration pulse sequences, see Ref.~\cite{Neill2018}.

\item Calibrate the distortion of the fast Z pulse.  Distortion information is obtained by probing the time-domain response of the qubit phase right after a Z pulse~\cite{Barends2014nature}.  We note that the distortion of the fast Z pulse  for the coupler is derived with the help of the phase response of its adjacent qubits. 
\end{enumerate}

With the information above,  we can start to tune up the universal quantum gate set on the $3\times6$ lattice, which includes 18 single-qubit rotations and 27 two-qubit CZ gates. This is challenging due to the existence of pulse distortions, TLS defects, and crosstalk.  We allocate a set of idle frequencies $\{\omega_{10}\}$ to qubits, which are optimized to yield high-fidelity single-qubit gates as well as to favor the implementation of two-qubit CZ gates.  We consider several important principles, which are listed below. 
\begin{itemize}
\item Energy relaxation time $T_1$ and spin-echo pure dephasing time $T_2^{\rm SE}$  in the vicinity of $\omega_{10}$ should be long and stable.
\item Frequency detuning between two qubits with stray coupling should be much larger than the strength of the stray coupling.
\item The fast Z-pulse amplitude for realizing a two-qubit CZ gate should be small to minimize the impact of residue pulse distortion. 
\end{itemize}

\begin{figure*}[h]
\center
\includegraphics[width=1.0\linewidth]{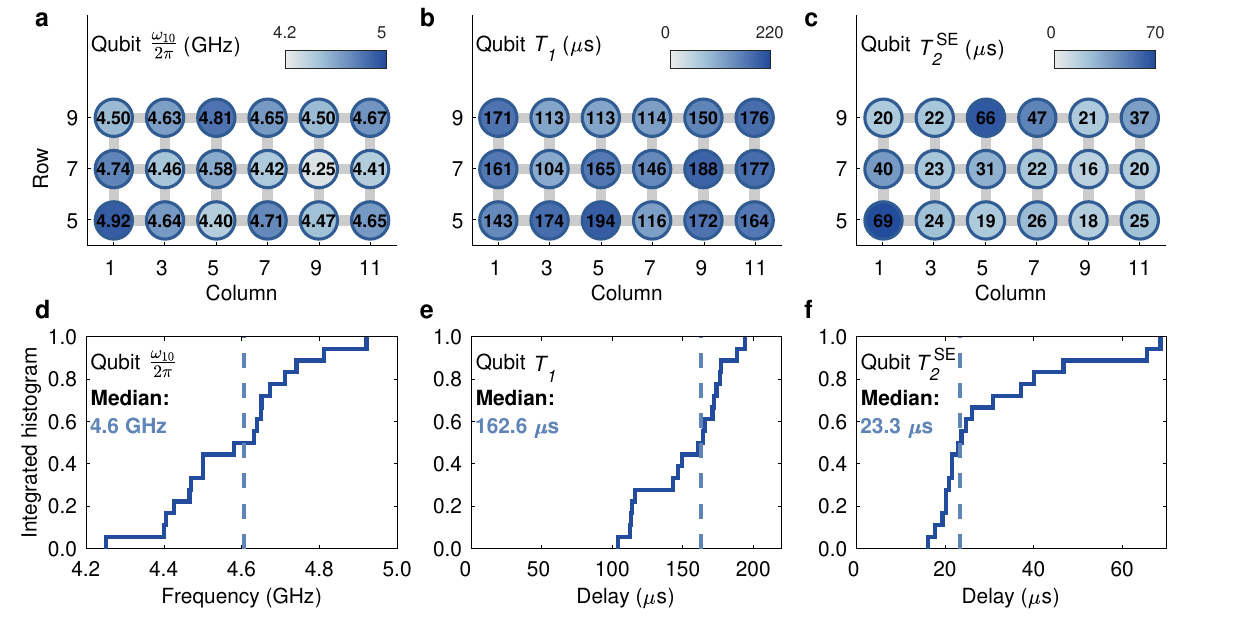}
\caption{\textbf{Qubit performance at idle frequencies.} 
{\bf a}, Heat map of  idle frequencies $\{\omega_{10}\}$. 
{\bf b}, Heat map of  qubit energy relaxation times $\{T_{1}\}$ at idle frequencies. 
{\bf c}, Heat map of qubit spin-echo pure dephasing times $\{T^{\rm SE}_{2}\}$ at idle frequencies.  {\bf d}, {\bf e}, and {\bf f} are the integrated histograms of $\{\omega_{10}\}$, $\{T_{1}\}$  and $\{T^{\rm SE}_{2}\}$, respectively, which are obtained using the data in {\bf a}, {\bf b}, and {\bf c}.} 
\label{fig:qubit_f10_T1_T2}
\end{figure*}

\begin{figure*}[h]
\center
\includegraphics[width=1.0\linewidth]{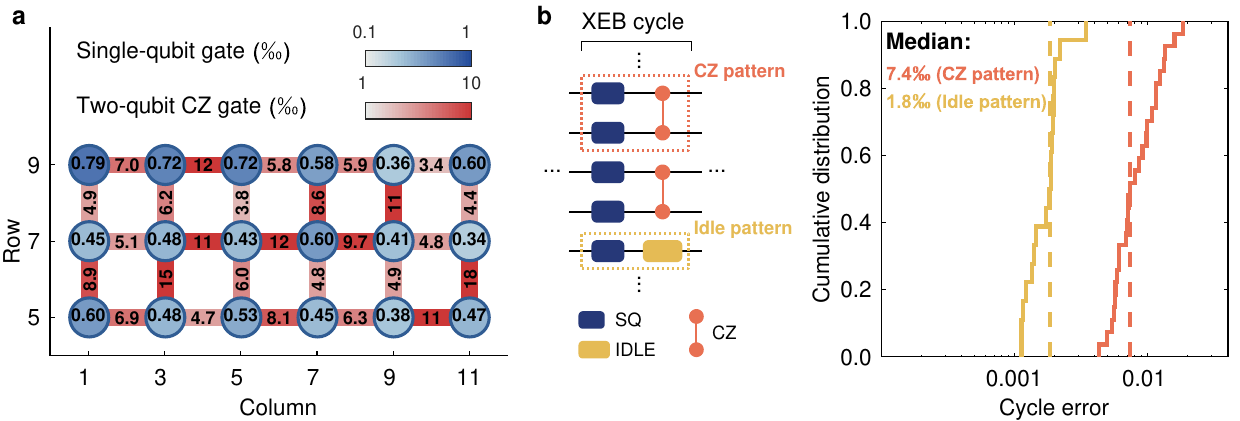}
\caption{\textbf{Gate errors.} 
%\textbf{a}, Heat map of Pauli errors for single-qubit gates (blue) and two-qubit CZ gates (red). The values are measured by performing XEB measurement for each gate individually.
{\bf a}, Heat map of layer-averaged Pauli errors for single-qubit gates (blue) and two-qubit CZ gates (red). These values are measured by performing simultaneous XEB using the typical single-qubit gate layers (SQ layers) and two-qubit CZ gate layers (CZ layers) in the Floquet unitary circuit at $B=0$ and in the eigenstate preparation circuit. Note that, for the simultaneous XEB of each CZ layer, those qubits that are not involved in CZ gates undergo a single-qubit XEB sequence.
{\bf b}, Schematic of the XEB circuit for CZ layers and cumulative distributions of the cycle error. The left panel shows the XEB circuit for the CZ layer. Each cycle is composed of a  single-qubit gate layer and a subsequent CZ layer, which includes two types of patterns, a CZ pattern and an idle pattern. The right panel shows the cumulative distributions of cycle errors for the CZ pattern~(red) and the idle pattern~(yellow). Each data point represents a specific pattern for the target gate, which is averaged over all CZ layers. Using cycle errors, we can estimate CZ gate errors in {\bf a} and idle gate errors in the CZ layers.
}
\label{fig:sq_and_cz_gate_error}
\end{figure*}

In each optimization round, we tune up all the single- and two-qubit gates,  and then test  their fidelities by performing simultaneous cross-entropy benchmarking (XEB)~\cite{Boixo2018NP} using the typical layers in the target circuits. These results  are used as the feedback for the next round of optimization. After several rounds, we obtain a set of idle frequencies and gate parameters for our experiments. The idle frequencies $\{\omega_{10}\}$ in this experiment are shown in Fig.~\ref{fig:qubit_f10_T1_T2}{a}.  The measured energy relaxation time $T_1$ and spin-echo dephasing time $T_2^{\rm{SE}}$  at  $\{\omega_{10}\}$  are listed in Fig.~\ref{fig:qubit_f10_T1_T2}{b} and {c}, respectively.  Their cumulative distributions and median values are shown in Fig.~\ref{fig:qubit_f10_T1_T2}{d}, {e}, and {f}.  Remarkably, the median value of $T_1$ over 18 qubits is $\sim$163~$\mu$s. We also achieve median Pauli errors ($\epsilon_p$) of $\sim$$0.48\times 10^{-3}$ for single-qubit gates and $\sim$$6.4  \times 10^{-3}$ for two-qubit CZ gates,  which is equivalent to the randomized-benchmarking fidelities~\cite{Google2019Supremaecy} of 0.9997 and 0.9949, respectively. Here we use the relation $F = 1 - \epsilon_p / (1 + 1 / 2^d )$, where $d$  is the number of qubits. Figure \ref{fig:sq_and_cz_gate_error} shows detailed information on gate errors.
% results for  gate error information under different conditions.

% The readout fidelities of qubits are optimized by preparing 18 qubits in random Fock product  states~\cite{Nation2021} and simultaneously measuring the corresponding readout fidelities for different control parameters.  

\begin{figure*}[h]
\center
\includegraphics[width=1.0\linewidth]{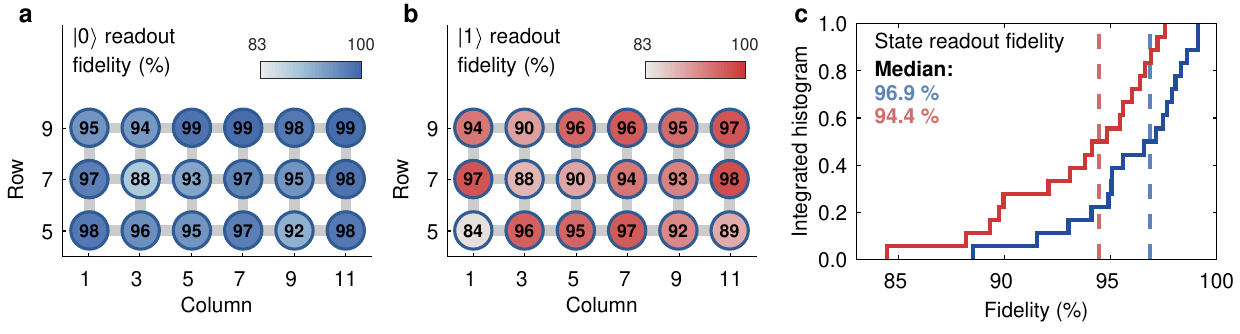}
\caption{\textbf{Qubit readout fidelity.} 
{\bf a}, Heat map of the readout fidelity for state $|0\rangle$.
{\bf b}, Heat map of  the  readout fidelity for state $|1\rangle$.  
{\bf c}, Integrated histogram of 18 readout fidelities based on the data in {\bf a} and {\bf b}.  To obtain the readout fidelities, we prepare all the qubits in random computational-basis product states and perform simultaneous measurements. For each qubit,  its  $|0\rangle$ ($|1\rangle$) readout fidelity is given by calculating the $|0\rangle$ ($|1\rangle$)  state probability from the samples that this qubit is prepared in  $|0\rangle$ ($|1\rangle$).}
\label{fig:readoutFidelities}
\end{figure*}

Readout fidelities of qubits are simultaneously measured by preparing 18 qubits in random  product states~\cite{Nation2021} and averaging them for each qubit.  %We optimize the parameters of readout pulses and qubit frequencies to reduce the crosstalk during the measurement process. 
Fig.~\ref{fig:readoutFidelities} displays the measured readout fidelities for each qubit in our experiment,  which are also used to correct the effects of readout errors.

\subsection{Microwave  crosstalk}
Single-qubit rotations for each qubit are realized by applying microwave pulses to its microwave drive line. However,  %due to the spatial nonlocality,  
microwave pulses applied to $Q_i$ will also induce unwanted state transitions of $Q_j$. The microwave pulse crosstalk felt by $Q_j$ can be modeled as $\tilde{\Omega}_j = A_{ji}e^{-i\phi_{ji}}\Omega_i$~\cite{Sung2021}, where $\Omega_i$ is the microwave signal applied to the source qubit $Q_i$, while $A_{ji}e^{-i\phi_{ji}}$ describes the relative amplitude and the relative phase of the effect on the target qubit $Q_j$. It can be canceled by actively applying an opposite signal $-\tilde{\Omega}_j$ to the qubit $Q_j$. We use randomized benchmarking to detect microwave crosstalk, and use the measured matrix $\{A_{ji}e^{-i\phi_{ji}}\}$  to suppress such effects. 

\subsection{Flux-bias crosstalk}
\label{secs:z-crosstalk}
\begin{figure*}[bt]
\center
\includegraphics[width=1.0\linewidth]{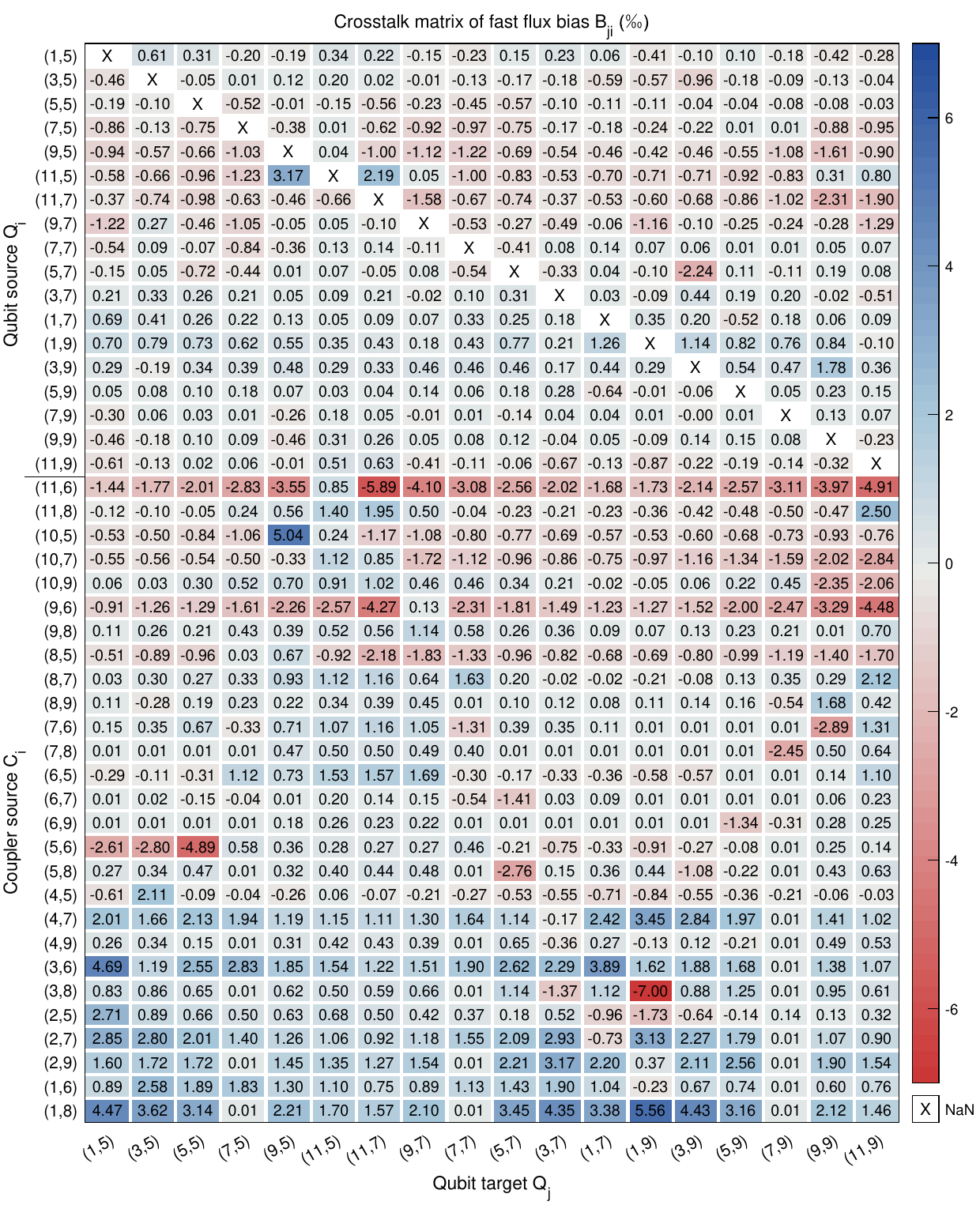}
\caption{\textbf{Crosstalk matrix of the fast flux bias.}  Each qubit (coupler) is labeled by $(x, y)$,  which means it is located in the $x$-th column and $y$-th row in our device.  Note that we neglect crosstalk  to couplers in our experiments. }
\label{fig:zxtalk}
\end{figure*}

In our device, applying a bias current to the flux line of qubit $Q_i$ (or coupler $C_i$) can cause a nonzero flux on other qubits (or couplers). During parallel gate operations, this flux crosstalk can introduce extra phase errors into the circuit.  The crosstalk %signal at the target qubit $Q_j$ from the source qubit $Q_i$ (or coupler $C_i$) can be simply 
can be modeled as $\tilde{\Delta}_j = \Delta_i B_{ji}$, where $\Delta_i$ is the flux to qubit $Q_i$ (or coupler $C_i$), $\tilde{\Delta}_j$ is the crosstalk flux felt by qubit $Q_j$, and $B_{ji}$ is the crosstalk ratio.  We neglect crosstalk  to couplers  in our experiments. To compensate for $\tilde{\Delta}_j$, we measure the ratio $B_{ji}$ and apply a flux bias $-\tilde{\Delta}_j$ to  $Q_j$. The measured crosstalk matrix elements $\{B_{ji}\}$ for fast Z bias are shown in Fig. \ref{fig:zxtalk}.  

\subsection{Device-aware circuit transformation}
\label{secs:circuit_transformation}
\begin{figure*}[h]
\center
\includegraphics[width=1.0\linewidth]{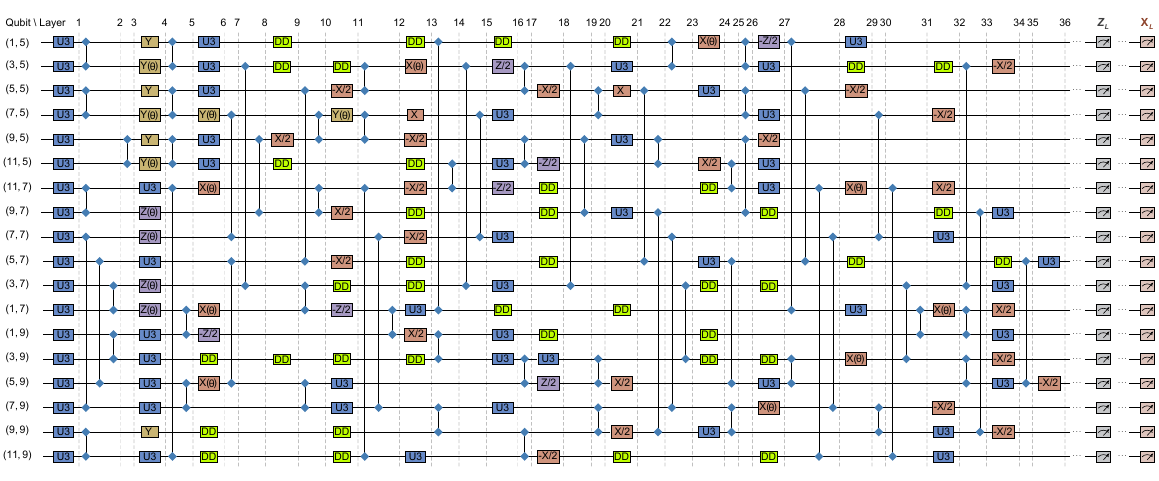}
\caption{\textbf{Experimental circuit to implement the Floquet unitary $U_{F}$ at $B=0.1$.}  The first layer of the circuit encompasses the spin-flip driving designed for $H_1$, while the subsequent layers of the circuit simulate the evolution of the rotated surface-code Hamiltonian $H_2$. $U_F$ is composed of 15 SQ layers and 21 CZ layers. In total, it contains 117 single-qubit gates (excluding 40 dynamical decoupling (DD) gates) and 71 two-qubit CZ gates. We have three types of single-qubit gates: Clifford gates \{$X$, $Y$, $Z$, $\pm X/2, \pm Y/2, \pm Z/2$ \}, parameterized rotation gates $\{X(\theta)$, $Y(\theta)$, $Z(\theta)\}$, and $\{ U3(\alpha, \theta,\phi)\}$ gates. The DD gates are inserted in pairs. In our experiments, this circuit is repeated up to 20 cycles, for a total of about 2340 single-qubit gates and 1420 CZ gates. This equates to a time-domain sequence of $\sim$28.8 $\mu s$. Note that, at $B=0$, single-qubit gates in the $H_1$ circuit are Clifford gates, and thus the whole circuit can be further simplified by merging some of them into the $H_2$ circuit.}
\label{fig:cir_u1u2}
\end{figure*}

A quantum circuit constructed theoretically usually incorporates little information about the limitations or imperfections of the underlying hardware performance, leaving space for further improvements before it is converted to physical control pulses. Therefore, it is worthwhile to transform the circuits generated in Section~\ref{secs:theory} with the awareness of the device information to further improve the circuit fidelity. Figure~\ref{fig:cir_u1u2} shows the transformed circuit of the Floquet-evolution unitary for a single period $T$ at $B=0.1$. We summarize the strategies and tools we use in this process below.

\begin{enumerate}
\item Remove redundant Clifford gates using ZX-calculus~\cite{kissinger2020Pyzx, Kissinger2020NonCliffordZXcalc}.
% The Floquet unitary circuit is composed of both Clifford gates and parameterized non-Clifford gates. After this step, the circuit structure is intact, i.e. only the parameters $\theta_p$ change for random circuit instances.
\item Use Qiskit~\cite{Qiskit} to convert the circuit into combinations of  two-qubit CZ gates and single-qubit rotations around the $x,y,z$-axis on the Bloch sphere and Cirq~\cite{Cirq} to identify single-qubit layers and separate them from CZ gates. Then we get a circuit %whose structure %is %reminiscent of a sandwich cookie:  
that alternates between layers of single-qubit and CZ gates: a single-qubit gate layer (SQ layer), followed by a layer of CZ gates (CZ layer), followed by an SQ layer, etc. 
\item Compile consecutive SQ gates into a U3 gate. The U3 gate is constructed by combining a $\theta$-angle rotation around the $z$-axis followed by an $\alpha$-angle rotation around an axis in the $xy$ plane with an azimuthal angle $\phi$, which can be written in the form

\begin{equation}\label{eqs:define_U3}
\begin{array}{cc}
\mathrm{U3}(\alpha, \phi, \theta) = {R_{xy}}(\alpha, \phi){R_{z}}(\theta) = \left[
\begin{array}{cc}
\cos \left(\frac{\alpha}{2}\right) & -i e^{-i \phi} e^{i \theta} \sin \left(\frac{\alpha}{2}\right) \\
-i e^{i \phi} \sin \left(\frac{\alpha}{2}\right) & e^{i \theta} \cos \left(\frac{\alpha}{2}\right)
\end{array}
\right]
\end{array}.
\end{equation}

Note that $R_{z}(\theta)$ is implemented virtually by adding an extra $\theta$ to the phase of the subsequent microwave pulse~\cite{McKay2017VZgate}. $R_{xy}(\alpha, \phi)$ is implemented by applying a microwave pulse, whose phase is $\phi$ and whose amplitude depends on  the rotation angle  $\alpha$.
\item Separate CZ gates in a given layer into several groups. This step is to avoid leakage caused by qubit level crossings while multiple CZ gates operate in parallel. A maximum of two groups are enough for our Floquet-evolution circuit. 
\item In the eigenstate circuit,  we align the gates to the right of the circuit  to delay the first operation on the qubit. Additionally, for quantum state tomography measurements, tomographic rotation is combined with an SQ gate at the end of the circuit, resulting in a U3 gate. To mitigate qubit-dephasing effects, these U3 gates are aligned to the left prior to measurement.
\item 
To suppress dephasing errors during idling, we incorporate dynamical decoupling (DD) gates. These DD gates are inserted within the circuit,  which effectively suppresses the dephasing and thus enhances the overall performance. 
\end{enumerate}

\section{Numerical simulations}

For the experimentally-studied system size, we can numerically simulate the dynamics determined by specifically designed quantum circuits in our experiment. First, classical simulations allow us to evaluate  the feasibility of the theoretical proposal by examining whether observables remain discernible after many Floquet cycles under realistic, noisy quantum gates. Second, we can deepen our understanding of our device performance by comparing the experimental results with simulation predictions that incorporate error models.

\subsection{Error model and noisy simulation}
\label{sec:sim_method}
We employ the Monte Carlo wavefunction method ~\cite{Mlmer1993MonteCarlo} to numerically simulate noisy circuits. The idea is to sample error operators according to a noise model and randomly insert them after each ideal gate. In this way, errors occur randomly in the circuit. We evolve the state vector along many quantum trajectories corresponding to many noise realizations. To obtain the value of the desired observable, we average over an ensemble of quantum trajectories, which resembles repeated measurements for evaluating the expectation value of an observable in real experiments. Notably, the Monte Carlo wavefunction method requires fewer computational resources compared to the master-equation approach, because it only stores state vectors of size  $2^{N}$ during the calculation rather than density matrices of size $2^{N}\times2^{N}$. In this context, we use the state-vector simulator provided by Qiskit for the numerical calculation of system dynamics. Qiskit provides several APIs to construct noise models that approximate the behavior of noisy circuits executed on real NISQ devices. 

We model realistic errors with different quantum channels, and represent them with a probabilistic mixture of different operators. The corresponding parameters are estimated using the experimental benchmarks of gate errors and device performance. The following provides an introduction to the error model.
\begin{enumerate}
    \item {\bf Decoherence errors.} 
Due to interactions with the environment, energy relaxation and dephasing naturally occur in the dynamics can be described by the quantum channel
\begin{equation}
\label{eq:qubit_decay_dephasing}
\mathcal{E}(\rho)=\left(\begin{array}{cc}
        1-\rho_{11} e^{-t / T_1} & \rho_{01} e^{-t / T_2} \\
        \rho_{10} e^{-t / T_2} & \rho_{11} e^{-t / T_1}
        \end{array}\right) = \sum^3_{i=0} M_i\rho M^\dagger_i,
\end{equation}
where $\rho$ is the density matrix of a single qubit with elements $\rho_{ij}$ ($i,j=0,1$).
Here, $T_1$ represents the energy-relaxation time originating from energy exchange with the environment, and $T_2$ denotes the dephasing time characterizing the damping of off-diagonal terms of the density matrix. They satisfy the relation
\begin{equation}
 \frac{1}{T_2}=\frac{1}{2 T_1}+\frac{1}{T_\phi},
\end{equation}
where $T_\phi$ is the pure dephasing time, arising from non-dissipative interactions with the environment. This quantum channel can be written in terms of the Kraus operators
\begin{equation}
\begin{array}{l}
M_0 = \sqrt{1-p_0-p_1} (|0\rangle\langle0| + |1\rangle\langle1|), \\
M_1 = \sqrt{p_0} |0\rangle\langle0|, \\
M_2 = \sqrt{p_0} |0\rangle\langle1|, \\
M_3 = \sqrt{p_1} (|0\rangle\langle0| - |1\rangle\langle1|),
\end{array}
\end{equation}
which satisfy the normalization condition $\sum^3_{i=0}M^\dagger_i M_i=I$.
The Kraus operator $M_0\propto I$ indicates that the qubit remains intact with probability $1-p_0-p_1$. The pair of Kraus operators $M_1$ and $M_2$ describe spontaneous decay of the qubit from its excited state $|1\rangle$ to its ground state  $|0\rangle$. This is realized in our simulations by randomly applying a reset operation to the qubit with probability $p_0=1 - e^{-t/T_1}$.  $M_3$ contributes an additional dephasing channel to the qubit. Together with the phase damping caused by $M_1$ and $M_2$, $M_{1,2,3}$ describe the total phase damping.  $M_3$ is realized in our simulations by randomly applying $\sigma^z$ operators with probability $p_1=\frac{1}{2}e^{-t/T_1}[1 - e^{-t(1/T_2-1/T_1)}]$.  The two simulation parameters $p_0$ and $p_1$ are estimated using the average values of $T_1$ and $T_2^{\rm{SE}}$ reported in Section~\ref{secs:chip_characterization}, while the value of $t$ is set by the averaged time required to apply an SQ (CZ) layer. 

    \item {\bf Depolarizing errors.} The depolarizing channel is defined as
\begin{equation}
\mathcal{E}(\rho)=\left(1-e_p\right) \rho+\frac{e_p}{4^d-1} \sum_{\mu \neq 0} P_\mu \rho P_\mu,
\end{equation}
where $d$ is the number of qubits, $\rho$ is a $d$-qubit density operator, $P_\mu \in \{I, X, Y, Z\}^{\otimes {d}}$ is the tensor product of Pauli gates, and $e_p$ denotes the Pauli error per cycle.   We use the depolarizing channel to account for errors caused by imperfect control of the system, such as gate control errors and crosstalk errors. It is realized in our simulations by applying a randomly chosen non-identity Pauli string with probability $e_p/(4^d-1)$. 
\end{enumerate}

\label{secs:error_budget}
Using the error model above, we perform numerical simulations to verify the observed results in the main text. The decoherence error for each SQ (CZ) layer, quantified by $p_0$ and $p_1$ in the model, is estimated with $T_1$ and $T_2$ fixed by the average measured $T_1$, $T_2^{\rm{SE}}$ values (see Fig.~\ref{fig:qubit_f10_T1_T2}), and with $t$ set by the pulse duration corresponding to an SQ or CZ layer, depending on whether the error channel is being applied after a single- or two-qubit gate layer. The average pulse duration is about 24.0 ns for an SQ layer and 62.6 ns (52.5 ns) for a CZ layer in circuits with (without) eigenstate preparation.
Then, the depolarization error rate $e_p$ for each type of gate is estimated by subtracting the decoherence error rate from the median Pauli error rate $\epsilon_p$ estimated in the previous section~(Section~\ref{SystemCalibration}). Here, Pauli errors are characterized using XEB experiments (see Fig.~\ref{fig:sq_and_cz_gate_error}), whose median value is $\epsilon_p\sim0.48 \times 10^{-3}$ for a single-qubit gate, $0.64 \times 10^{-2}$ for qubits involved in a CZ gate, and $1.37 \times 10^{-3}$ ($1.10 \times 10^{-3}$) for qubits that are idle during a CZ layer, in experimental circuits with (without) eigenstate preparation.
Simulation results using the error sources above are shown in Fig. 2{a} and Fig. 4{b}, {c}, and {e} of the main text and exhibit good agreement with the experiments.

\bibliography{SI_refs}